\ifx\XeTeXversion\undefined\pdfoutput=1\fi 
\documentclass[11pt]{article}

\usepackage{iftex}
\ifPDFTeX
  \usepackage[T1]{fontenc}
  \usepackage[utf8]{inputenc}
\fi
\usepackage{amsmath}
\usepackage{newtxtext,newtxmath}
\usepackage[margin=1.1in]{geometry}
\usepackage{booktabs}
\usepackage{array}
\usepackage{graphicx}
\usepackage{xcolor}
\usepackage[font=small,labelfont=bf]{caption}
\usepackage{enumitem}
\usepackage[authoryear,round]{natbib}
\usepackage[colorlinks=true,linkcolor=blue!50!black,citecolor=blue!50!black,
            urlcolor=blue!50!black]{hyperref}
\usepackage{microtype}
\usepackage{float}

\setlist{itemsep=2pt,topsep=4pt}
\newcolumntype{L}[1]{>{\raggedright\arraybackslash}p{#1}}
\newcommand{\ci}[2]{{\small$[#1,#2]$}}

\title{Conformal Kelly: When Marginal Coverage Transfers\\[2pt]
and the Economics Do Not\thanks{Extended version of the paper of the same
title to appear in the Proceedings of the 7th ACM International Conference on
AI in Finance (ICAIF '26), Milan, Italy, November 14--17, 2026. Version~1 of
this preprint (August 2026) was titled \emph{Conformal Kelly: Conformal
Prediction Intervals as the Scale in Fractional Kelly Position Sizing};
Section~\ref{sec:v1} lists what changed. Version~2 (October 2026) was titled
\emph{When Marginal Coverage Transfers and the Economics Do Not: A
Pre-Registered Study of Conformal Interval Widths as Position-Sizing Scales};
version~3 adopts the conference title and updates the affiliation; the content is unchanged.}}
\author{\textbf{Robert Jacob Ryan}\\ Independent Researcher, Athens, Greece\\
ORCID: \href{https://orcid.org/0009-0000-9458-9095}{0009-0000-9458-9095}}
\date{October 2026}

\begin{document}
\maketitle

\begin{abstract}
\noindent
We size portfolio positions from conformal prediction intervals, reading the
interval half-width as the scale in a capped fractional-Kelly rule
(\emph{Conformal Kelly}), and evaluate the rule under a pre-registered
protocol. Its development record is an autonomous LLM-agent search over
roughly 200 configurations on a single six-year window (2016--2021) of eight
US-listed ETFs. All later data was sealed before the search began, and two
final configurations, nine comparison books, ten metrics, a written growth
prediction and the mapping from outcomes to conclusions were registered before
unsealing.

On the sealed 2022--2024 window, realized conformal coverage was 0.745 against
a 0.750 target, but the two configurations' annualised net log growth was
8.5\% and 7.0\% (28.5\% and 25.8\% in development), below an unlevered
equal-weight book (9.5\%), the leveraged passive books (15.6--16.8\%) and a
registered hindsight four-equity book (10.8\%, at a 46\% drawdown). Both
configurations ranked last of eleven entries on Sharpe and Calmar, and the
drawdown control failed its registered criterion. A post-hoc split, outside
the registered protocol, shows the passing coverage averaging 0.592 in 2022
and 0.842 afterwards, with the entire loss in the under-covered year. The
split does not show that under-coverage caused the loss: the leverage cap
bound on every 2022 trading day, so narrow intervals could not enlarge the
book, and a long-equity book in an equity bear market is the simpler account.
But the registered acceptance test passed a sixteen-point conditional failure,
which motivates, without establishing, conditional coverage as an endpoint for
rules that read interval width as a scale.

On the development window, a slow, unweighted, per-asset conformal quantile
out-grew every locally adaptive alternative we tested (adaptive conformal
inference, recency-weighted and asymmetric intervals, pooled and Mondrian
calibration) and a plug-in standard deviation by 2.1 percentage points a year
at matched leverage. That evidence is exploratory and cap-conditional: it was
measured under a gross-leverage cap binding on 97.7\% of days, and
dependence-aware block-bootstrap intervals include zero for 11 of the 13
contrasts we could re-derive.
\end{abstract}

\section{Introduction}
\label{sec:intro}

Two literatures have grown up next to each other without meeting. Conformal
prediction turns a point forecaster into an interval forecaster with a
finite-sample, distribution-free \emph{marginal} coverage
guarantee~\citep{vovk2005}; the Kelly criterion turns a belief about a return
distribution into the bet size that maximises expected log
wealth~\citep{kelly1956}. The first produces something close to what the
second needs: a calibrated statement of how wrong a forecast tends to be. The
match is inexact, because Kelly consumes a full return distribution while an
interval supplies a single scale (Section~\ref{sec:method}). To our knowledge,
no prior work \emph{empirically studies} conformal intervals as the scale input
to Kelly sizing (Section~\ref{sec:related}).

Kelly needs a \emph{scale}: a $\sigma$ such that $f^\star\approx\mu/\sigma^2$
is the right fraction of wealth to commit. Reading $\sigma$ off a conformal
half-width, the rule we call \emph{Conformal Kelly}, is trivial to write down,
and it changes what a good interval is. A forecaster wants an interval sharply
adapted to local conditions; a sizing rule may not, because a noisy $\sigma$
enters the sizing map nonlinearly and its estimation variance is charged to
compounded wealth. And a marginal guarantee constrains how often an interval
fails on average, not when it fails.

This paper reports a negative result obtained under a pre-registered protocol,
and it orders its claims by the strength of the evidence behind them. The
empirical base is one universe of eight exchange-traded funds (ETFs), one
six-year development window and one 2.7-year sealed window; every general
statement below should be read against that single experiment.

\textbf{(i) Confirmatory: a pre-registration protocol for agentic search, and
its outcome.} The configuration space was searched by an autonomous LLM agent
against an immutable evaluation harness (Section~\ref{sec:agentic},
Appendix~\ref{app:ledger}), a large multiple-comparisons
exposure~\citep{bailey2014dsr,bailey2017pbo} that we report rather than omit.
The mitigation is structural: a window sealed before the search began, and a
registration, committed before unsealing, that fixed the two final
configurations by commit hash, ten metrics, nine comparison books including the
one most likely to beat ours, a sample-size caveat, a written prediction, and
the mapping from outcomes to conclusions. Under that mapping the outcome is a
partial refutation: coverage transferred, growth did not, the registered
risk-adjusted defence failed, and so did the drawdown dial's registered
criterion (Section~\ref{sec:lockbox}).

\textbf{(ii) Post hoc: a conditional-coverage diagnostic.} Splitting the
sealed window at the year boundary shows realized coverage of 0.592 in 2022
and 0.842 in 2023--24: the passing marginal figure averages two large errors of
opposite sign, and the under-covered year holds the entire loss. The split was
not registered, so the paper's most informative observation lies outside its
confirmatory protocol; we state that tension wherever the observation is used
and do not resolve it. Nor does the split identify a mechanism. The leverage
cap bound on every trading day of 2022, and a common cause---a volatility
regime that outran a deliberately slow estimator while equities fell---fits
the data at least as well as a causal link (Section~\ref{sec:split}). What
the split does show is that the registration's own acceptance test passed a
window containing a sixteen-point conditional failure. That marginal validity
does not imply conditional validity is known~\citep{vovk2012,barber2021limits};
one documented instance of the gap in a sizing rule motivates, but does not
establish, conditional coverage as a registered endpoint for rules that read
an interval width as a scale.

\textbf{(iii) Exploratory and cap-conditional: evidence on conformal scale
estimation.} On the development window, under a gross-leverage cap that bound
on 97.7\% of days, every device we tested that made the interval adapt faster
lost growth, and the conformal quantile out-grew a plug-in standard deviation
by 2.1 percentage points (pp) a year at matched leverage
(Sections~\ref{sec:dev} and~\ref{sec:refute}). A drawdown control built from
the same intervals cut development-window drawdown at better Sharpe, surviving
a constant-leverage control and a timing placebo (Section~\ref{sec:dial}). The
direction of the scale evidence is consistent across devices, but the entries
were selected from a searched grid, block-bootstrap intervals include zero for
11 of the 13 contrasts we could re-derive, and off the cap the ordering is
untested (Section~\ref{sec:limits}).

\subsection{What changed from version 1}
\label{sec:v1}

This version reorganizes the paper around its registered out-of-sample result
and retitles it; version~1 led with development-window performance. It adds
(a) a post-hoc split of the sealed window by year, with a check of its
mechanism against the paper's own leverage-cap fact (Section~\ref{sec:split});
(b) moving-block-bootstrap intervals, computed after unsealing on re-derived
series of configurations scored before it, for the development contrasts and
for sealed-window coverage and comparisons (Appendix~\ref{app:postreg}), which
leave most development contrasts within noise; and (c) an attribution of the
fall in growth from development to the sealed window
(Section~\ref{sec:attrib}). Claims about the development window are restated
with that uncertainty, and mechanisms that were not measured are labelled as
candidates. Two descriptions are corrected: the primary and secondary variants
differ on 13 trading days at the end of the sample, not on ``the final 21
rows''; and the statement that the conformal machinery transferred through
2022 holds only for coverage pooled over the whole window. No
development-window number and no registered sealed-window number has changed.

\section{Related work}
\label{sec:related}

\textbf{Conformal prediction.} The split-conformal construction and its
finite-sample marginal coverage guarantee are due to \citet{vovk2005}.
Financial series violate exchangeability, and the field's standard responses
are the ones we test and reject on the development window
(Section~\ref{sec:refute}): adaptive conformal inference~\citep{gibbs2021},
which updates $\alpha$ online from realized coverage errors; non-exchangeable,
recency-weighted conformal prediction~\citep{barber2023}, which downweights
stale calibration points; and asymmetric intervals in the spirit of
conformalized quantile regression \citep[CQR;][]{romano2019}.
EnbPI~\citep{xu2021}, the ensemble-bootstrap construction for dependent
series, is related but not tested here. Mondrian conformal
prediction~\citep{vovk2003} supplies the within-taxonomy pooling of
Section~\ref{sec:refute}. Conformal predictive systems~\citep{vovk2019} and
their decision-theoretic use~\citep{vovk2018}---expected-utility maximisation
over a predictive distribution, which under log utility \emph{is} Kelly---are
the theoretically correct version of what we do. We implement the decision rule
(expected-log-utility maximisation over the empirical calibration-residual
distribution) rather than a full conformal predictive system, and it loses,
for two measurable reasons (Section~\ref{sec:cps}). Exact conditional coverage
cannot be guaranteed distribution-free~\citep{vovk2012,barber2021limits};
recent methods guarantee coverage over prespecified subgroups or classes of
covariate shift~\citep{gibbs2025}, the natural tool for the conditional
endpoint that Section~\ref{sec:split} proposes. \citet{noguer2024b} surveys
conformal prediction in finance.

\textbf{Growth-optimal betting.} \citet{kelly1956} and the fractional- and
drawdown-constrained Kelly literature
\citep{maclean1992,grossman1993,browne1997,maclean2010,maclean2011} supply the
sizing map and the reason to use a \emph{fraction} of full Kelly: its drawdown
behaviour is intolerable and its sensitivity to estimation error in $\mu$ and
$\sigma$ severe. Our object is orthogonal: not a new sizing map, but an
estimator for its denominator.

\textbf{The intersection.} Three strands come close, on different axes.
\citet{kato2024} is the closest \emph{selection} work: he conformalises
per-portfolio return forecasts and applies a two-stage screen---keep the top
$m$ candidates by the interval's \emph{lower} bound, then take the argmax of
the \emph{upper} bound---over a \textbf{finite candidate set} of weight vectors,
on three-asset universes of monthly returns. Both endpoints are used, but only
as a ranking statistic over a discrete menu: no continuous allocation, no
leverage decision, no growth objective. \textbf{He selects; we size.} On the
\emph{allocation} axis, \citet{noguer2024} integrates conformal intervals into
mean--variance optimisation, and \citet{jiahan2026} derive a value-at-risk from
the conformal lower bound and optimise continuous weights against a
VaR-constrained objective by projected gradient descent: continuous weights,
but risk-based rather than growth-optimal objectives, and no leverage
multiplier. On the \emph{Kelly} axis, \citet{sunboyd2018} optimise Kelly
against an ambiguity set over the return distribution, the natural home for a
conformal construction, though they do not build one; and the ARTEMIS
preprint~\citep{ray2026} proposes, in an optional and never-evaluated appendix
module, approximating a Kelly objective's covariance from conformal intervals,
with no portfolio results reported. The words ``Kelly'', ``bet sizing'' and
``position sizing'' do not appear in \citet{kato2024}. Our decision variable is
a continuous fraction of wealth, our objective is $\mathbb{E}[\log W]$, and our
central object is the interval \emph{width}, evaluated end to end under costs
and a sealed out-of-sample protocol. To pre-empt a keyword collision: Vovk's
Bayes--Kelly work~\citep{vovk2024} uses Kelly betting inside conformal
\emph{testing}, not investment sizing. Our literature search covered arXiv,
SSRN and the major ML and finance proceedings through August 2026.

\textbf{Backtest overfitting.} The deflated Sharpe ratio~\citep{bailey2014dsr}
corrects one statistic for a known number of trials, and the probability of
backtest overfitting~\citep{bailey2017pbo} assesses a selection procedure. A
registration is complementary: it protects the \emph{interpretation}---which
comparators may be dropped, which outcomes promoted (Section~\ref{sec:lockbox}).

\textbf{Autonomous research loops.} The development methodology is the
fixed-harness agentic pattern of Karpathy's
\texttt{auto\-research}~\citep{karpathy2026}: an immutable \texttt{prepare.py}
defining data, splits, costs and the metric; a mutable \texttt{train.py} that
is the agent's entire action space; and a keep-or-revert loop on a held-out
score. The academic analogues are \emph{The AI Scientist}~\citep{lu2024} and
\emph{MLE-bench}~\citep{chan2024}. We are not aware of an earlier
quantitative-finance study that discloses an agentic search in its method
section rather than omitting it, though we have not surveyed for this
systematically (Section~\ref{sec:agentic}).

\section{Method}
\label{sec:method}

Fix an asset universe $\{1,\dots,A\}$ and let $r_{i,t}$ be the simple daily
return of asset $i$. Write $R^{(H)}_{i,t}=\sum_{s=1}^{H} r_{i,t+s}$ for the
forward $H$-day arithmetic return sum; it differs from the compounded simple
return at second order (Section~\ref{sec:robust}), and nothing below is a true
cumulative return.

\textbf{Conventions.} We take the risk-free rate as zero throughout:
$\hat\mu$ is a total-return forecast, Sharpe is computed on raw rather than
excess returns, and financing enters only as the ex-post overlay of
Section~\ref{sec:robust}. That is benign on the development window, where
2016--2021 short rates sat near zero, and materially flattering on the sealed
window, where 2022--2024 cash paid 4--5\% (Section~\ref{sec:limits}).

\textbf{Forecast.} For each asset independently, an expanding-window ridge
regression ($\lambda_{\text{ridge}}=10$, refit every 21 trading days, first
prediction after 750 days) maps four unstandardised features---momentum over
21, 63 and 252 days and an exponentially weighted 20-day volatility---to
$R^{(H)}$, with training truncated at \texttt{fit\_end = i - H + 1} so every
label has landed by the prediction date. The forecaster is deliberately weak:
heavy ridge shrinkage toward the intercept makes much of its output a slowly
varying drift estimate, and a \emph{drift-only} forecast (pure intercept, zero
features) already reaches 0.2326, 82\% of the final development result.
\textbf{The point forecast is not where this paper's contribution lives.}

\textbf{Horizon ensemble.} The forecast is computed at
$H\in\{12,16,21,27,34\}$, each component rescaled by $21/H$ to a common 21-day
reference and combined by \emph{conformal inverse-variance model averaging}:
component $h$ gets weight $1/q_h^2$, $q_h$ being the causal conformal
half-width of that component's residual against the common target. The
conformal object is thus used twice, as a model-averaging weight and as the
sizing denominator.

\textbf{Conformal interval.} The nonconformity score is the plain absolute
residual $s_{i,t}=|R^{(21)}_{i,t}-\hat\mu_{i,t}|$. At date $t$ the conformal
quantile is the $(1-\alpha)$ empirical quantile of the last $W=500$ scores
\emph{landed} by $t$ (their outcome is observed), over the window
$[t-H-W+1,\,t-H+1)$, so no score whose outcome is still in the future can
enter. We use $\alpha=0.25$, taking the plain $(1-\alpha)$ empirical quantile
rather than the finite-sample conformal index $\lceil(n+1)(1-\alpha)\rceil/n$:
at $W=500$ the two differ by less than one rank in five hundred, and since
exchangeability fails for overlapping financial returns in any case, coverage
in this paper is a measured quantity, not an inherited guarantee. The rolling
quantile is then geometrically shrunk toward an \emph{expanding} anchor,
\begin{equation}
q^{\text{eff}}_{i,t}=\bigl(q^{\text{roll}}_{i,t}\bigr)^{1-\lambda}
\bigl(q^{\text{anchor}}_{i,t}\bigr)^{\lambda},\qquad \lambda=0.3,
\end{equation}
where $q^{\text{anchor}}$ is the $(1-\alpha)$ quantile of \emph{all} landed
scores to date, recomputed every 21 rows and held stale in between so it is
never early. It is the slowest updating scheme we field short of freezing
outright (a fully frozen estimator, Section~\ref{sec:sigma}, updates less and
loses in point estimate), and exists to test the slow-scale idea without the
window-length artifact (\texttt{min\_train}${}+W$ pushes the first usable date
into the development window once $W$ exceeds about 600).

\textbf{This is a heuristic scale estimator built from conformal quantiles, not
a split-conformal predictor.} The geometric shrinkage, the rolling window, and
the conformal-derived ensemble weights each break the exchangeability argument,
as do the forecaster's periodic refits. We report realized coverage as an
empirical property and claim no finite-sample guarantee for $q^{\text{eff}}$.
The sizing map is likewise Kelly-inspired rather than Kelly-optimal: under fat
tails $q^{\text{eff}}/z$ is a robust scale proxy, not a standard-deviation
estimate (the Gaussian constant fixes units only), covariance is ignored
(Section~\ref{sec:refute}), and the caps of Section~\ref{sec:data} bind almost
always, making the rule a capped inverse-scale-squared sizing rather than
textbook multi-asset Kelly.

\textbf{Sizing.} Read a volatility off the interval,
$\hat\sigma_{i,t}=q^{\text{eff}}_{i,t}/z_{1-\alpha/2}$ with the fixed constant
$z_{1-\alpha/2}=1.2816$, and apply fractional Kelly,
\begin{equation}
f_{i,t}=\kappa\cdot\frac{\hat\mu_{i,t}}{\hat\sigma^2_{i,t}},\qquad \kappa=0.15,
\end{equation}
winsorised at $\pm0.75$ per asset. Under the harness's gross cap the scale
$\kappa$ is nearly inert (Section~\ref{sec:data}); it is the \emph{ratios}
$\hat\mu_i/\hat\sigma^2_i$ that matter.

\emph{A disclosed inconsistency.} Gaussian consistency at $\alpha=0.25$ would
require $z_{1-\alpha/2}=\Phi^{-1}(0.875)=1.1503$; the frozen configurations
carry $z_{1-\alpha/2}=1.2816=\Phi^{-1}(0.90)$, left over from an earlier
$\alpha=0.20$. Since $z$ enters only through the overall scale, which the gross
cap makes inert, the mismatch is equivalent to a 24\% change in $\kappa$ (the
ratio enters squared: $(1.2816/1.1503)^2=1.24$) and is bounded rather than
measured: the per-asset clip precedes gross renormalisation, so the rescaling
is not exactly uniform, but the $\kappa$ sweep is flat across changes larger
than 24\%, and the registration forbids a post-unsealing sensitivity rerun. We
record rather than correct it, because correcting it would change the frozen
configurations.

\textbf{Book construction.} The traded book is the equal average of the target
books formed on each of the last $K=5$ days---staggered overlapping portfolios
in the sense of \citet{jegadeesh1993}---rather than an exponential decay
(Section~\ref{sec:holding}).

\textbf{The drawdown dial (Config B only).} Let $d_t$ be the trailing 21-day
\textbf{pooled} rate of \emph{lower}-interval breaks, with each miscoverage
indicator shifted by $H$ to its landing date. Multiply the whole book by
\begin{equation}
m_t=\mathrm{clip}\!\left(1-\beta\,\frac{d_t-\alpha/2}{\alpha/2},\ 0.25,\
1\right),\qquad \beta=1.
\end{equation}
At $\beta=1$ this has no free sensitivity parameter: the multiplier is one
minus the fractional excess of downside miscoverage over the $\alpha/2$
reference. That reference is a convention, not a guarantee---the symmetric
interval controls total miscoverage $\alpha$ and promises nothing about its
downside share---so the dial is an empirical indicator against a historical
norm; the guaranteed version would calibrate a separate one-sided lower bound,
which the search did not build. The dial never touches the interval, so
conformal coverage is identical with and without it (0.7483 in both
configurations on the development window).

\section{Data and evaluation protocol}
\label{sec:data}

\textbf{Universe and data.} Eight liquid US-listed ETFs---SPY, QQQ, DIA, MDY
(equity), GLD, SLV, USO, DBC (commodity)---daily adjusted closes from a frozen
Kaggle snapshot~\citep{malik2024}, common sample 2006-05 to 2024-09-20.

\textbf{Splits, fixed before any experiment.} TRAIN to 2015-12-31 (includes
2008); \textbf{DEV 2016-01-01 to 2021-12-31}, 1{,}511 trading days, the only
window the search ever scored, spanning the 2018 fourth-quarter selloff and the
2020 crash; \textbf{the sealed window, 2022-01-01 onward}, opened once, after
registration (the code calls it the lockbox). Table~\ref{tab:labels} collects
the labels used throughout.

\begin{table}[t]
\centering
\caption{Labels used in this paper.}
\label{tab:labels}
\small
\begin{tabular}{@{}L{3.0cm}L{11.6cm}@{}}
\toprule
label & meaning \\
\midrule
TRAIN & 2006-05 to 2015-12-31; used only to fit the forecaster \\
DEV & 2016-01-01 to 2021-12-31 (1{,}511 trading days), the only window the
search ever scored \\
sealed window & 2022-01-01 to 2024-09-20 (683 trading days), opened once,
after registration; the code calls it the lockbox \\
leg & one of three sequential sessions of the agent's search; an ablation is
run against the best configuration of the leg in progress, its \emph{base} \\
Config A & the configuration with the highest DEV growth over all legs (git tag
\texttt{best-leg3}) \\
Config B & Config A plus the drawdown dial of Section~\ref{sec:method} (tag
\texttt{paper-config-drawdown}) \\
primary, secondary & the configurations exactly as frozen; and a pre-declared
variant that also forecasts before a label has landed (\texttt{predict\_tail}),
and so trades 13 more days at the end of the sample \\
B1--B9 & the nine registered comparison books (Tables~\ref{tab:devbars}
and~\ref{tab:l2}) \\
growth & annualised net log growth, $252\times$ the mean of $\log(1+r_t)$, net
of 5\,bps per unit of turnover \\
pp & percentage points of growth per year \\
gross & $\sum_i|w_i|$; \emph{pre-cap} before and \emph{post-cap} after the
harness caps it at 2.0 \\
coverage & share of valid asset-days whose realized 21-day return falls inside
the interval; nominal 0.75 \\
Calmar; Ulcer index & annualised arithmetic return over maximum drawdown;
root-mean-square drawdown \\
\bottomrule
\end{tabular}
\end{table}

\textbf{Harness.} \texttt{prepare.py} is immutable and byte-identical to the
scaffold commit (blob SHA
\texttt{f7ee2730\allowbreak 86f044e4\allowbreak d3f2fada\allowbreak eca7ebc4\allowbreak 7adc4037}).
It applies, in order: a $\pm0.75$ per-asset clip; renormalisation to gross
$\le 2.0$; a \textbf{one-day implementation lag}; and 5\,bps per unit of
turnover. The metric is annualised net log-wealth growth on DEV, where
\textbf{net means net of transaction costs only}: financing costs on gross
exposure above 1.0 sit outside the harness by design and are quantified
separately (Sections~\ref{sec:robust} and~\ref{sec:lbdetail}), so no headline
number in this paper is financing-inclusive. The metric is the Kelly objective
itself, which penalises over-betting without an added risk term. Throughout,
Sharpe and Calmar use the arithmetic annualised net return $252\,\bar r$
against a zero risk-free rate, while growth is $252\,\overline{\log(1+r)}$;
the two differ by the variance drag $\sigma^2/2$. The harness floors Calmar's
denominator at 0.05, which never binds here.

\textbf{The structural fact that conditions every result.} At the frozen
configuration, \emph{pre-cap} gross leverage averages 4.28 and exceeds the 2.0
cap on \textbf{97.7\% of DEV days} for Config A, the undialled book (the dial
lowers Config B to 3.61 and 81.1\%), with unconstrained Kelly-optimal gross
estimated at roughly $3.6\times$ the cap. The book therefore sits on the
monotonically increasing part of the growth-versus-exposure curve, pinned at
maximum gross essentially always. Three consequences follow:
\begin{enumerate}
\item $\kappa$, $z_{1-\alpha/2}$, and any uniform rescaling of $\hat\sigma$
are inert under the cap: \emph{the interval sets the book's composition, not
its size}. They change only how many assets pin at the per-asset cap.
$\alpha$ is not pure scale, since it reshapes the cross-section, but it is
empirically flat to within 1.3\,pp once leverage is controlled
(Section~\ref{sec:robust}).
\item Tilting the cross-section toward low-volatility assets should lose:
below the Kelly optimum, more portfolio volatility at the same Sharpe is more
growth. This is our leading explanation for why the literature's
most-recommended device, the volatility-scaled nonconformity score, fails here.
\item Comparison books must be restated at gross 2.0. The bars shipped in
\texttt{prepare.py} are at gross 1.0, and comparing a $2\times$-levered strategy
with a $1\times$-levered benchmark is not a comparison.
\end{enumerate}

\textbf{Comparison books, restated at the real cap} (all mechanical, all scored
by the identical harness), in Table~\ref{tab:devbars}.

\begin{table}[t]
\centering
\caption{Development window: comparison books under the harness, and the two
frozen configurations. The registered labels B1--B9 are those of the
pre-registration (Section~\ref{sec:lockbox}).}
\label{tab:devbars}
\small
\setlength{\tabcolsep}{4.5pt}
\begin{tabular}{@{}llcccccc@{}}
\toprule
 & book & growth & Sharpe & max dd & Calmar & ann.\ vol & gross \\
\midrule
B1 & SPY buy \& hold, as the harness expresses it & 0.1226 & 0.976 & 26.1\% & 0.505 & 13.5\% & 0.750 \\
 & SPY buy \& hold, uncapped (outside the harness) & 0.1594 & 0.976 & 33.7\% & 0.521 & 18.0\% & 1.000 \\
B2 & equal weight, gross 1 & 0.1189 & 0.841 & 32.5\% & 0.403 & 15.6\% & 1.000 \\
B3 & vol-target risk parity 10\% & 0.1063 & 1.078 & 18.1\% & 0.616 & 10.4\% & 1.013 \\
B4 & equal weight $2\times$ & 0.2125 & 0.841 & 56.5\% & 0.465 & 31.2\% & 2.000 \\
B5 & inverse-vol risk parity $2\times$ & 0.2225 & 0.971 & 52.4\% & 0.494 & 26.7\% & 2.000 \\
B6 & vol-target risk parity 30\% & 0.2074 & 1.047 & 38.8\% & 0.600 & 22.2\% & 1.917 \\
B7 & \textbf{4 equity ETFs @ 0.5 (post hoc)} & \textbf{0.2934} & 0.963 & 60.9\% & 0.605 & 38.2\% & 2.000 \\
B8 & 4 commodity ETFs @ 0.5 (post-hoc control) & 0.0892 & 0.430 & 68.7\% & 0.228 & 36.5\% & 2.000 \\
B9 & equity, QQQ at the per-asset cap (post hoc) & 0.3119 & 1.007 & 59.4\% & 0.653 & 38.5\% & 2.000 \\
\midrule
 & \textbf{Config A}, metric-best & \textbf{0.2845} & \textbf{1.336} & \textbf{27.7\%} & \textbf{1.127} & 23.4\% & 1.960 \\
 & \textbf{Config B}, drawdown dial & 0.2584 & \textbf{1.386} & \textbf{20.3\%} & \textbf{1.376} & 20.1\% & 1.833 \\
\bottomrule
\end{tabular}
\end{table}

\emph{Note on the shipped SPY bar.} The $\pm0.75$ per-asset clip silently
reduces the \texttt{SPY = 1.0} bar to a 0.75-notional position, as its 13.5\%
realized volatility confirms, so \textbf{the quoted ``SPY buy \& hold
= 0.1226'' is $0.75\times$ SPY}. True unlevered SPY on DEV compounds at 0.1594
at 18.0\% volatility and Sharpe 0.98; the harness cannot express a
1.0-notional single-name book at all. We use the honest figure throughout, and
flag it because it inflates every ``beats SPY'' statement in the development
notes by about 3.7\,pp.

\section{Development-window results}
\label{sec:dev}

Both frozen configurations were replayed from their git tags through the
harness by the diagnostics script and reproduce their recorded metric to six
decimal places (Table~\ref{tab:headline}).

\textbf{Status and uncertainty.} Every comparison in Sections~\ref{sec:dev}
to~\ref{sec:robust} is exploratory: the grids were searched, and the winning
entries are selected values rather than pre-specified hypotheses. All are
paired contrasts on one return path. Version~1 noted that paired standard
errors are far smaller than the marginal one but reported no intervals. After
unsealing, we re-derived each contrasted arm's daily series from the
experiment ledger---each reproduces its recorded metric---and computed circular
moving-block bootstrap intervals for the paired growth differences (63-day
blocks, three forecast horizons; 10{,}000 resamples; 21- and 126-day blocks
agree qualitatively; Appendix~\ref{app:postreg}). No configuration is scored
for the first time. Paired standard errors are indeed small against the
marginal one (about 1.4\,pp for the headline $\sigma$ contrast, against 8.2\,pp
for Config A's growth itself), but not small enough: of the 13 development
contrasts we could re-derive, 11 have 95\% intervals that include zero. The
intervals are not adjusted for the search, so if anything they flatter the
contrasts. Quantities quoted without an interval were not re-derived. Only the
sealed window (Section~\ref{sec:lockbox}) carries confirmatory weight.

\begin{table}[t]
\centering
\caption{Headline, DEV 2016--2021.}
\label{tab:headline}
\small
\begin{tabular}{@{}lcc@{}}
\toprule
 & Config A & Config B \\
\midrule
annualised net log growth & \textbf{0.2845} & 0.2584 \\
Sharpe & 1.336 & \textbf{1.386} \\
maximum drawdown & 27.68\% & \textbf{20.26\%} \\
Calmar & 1.127 & \textbf{1.376} \\
annualised volatility & 23.36\% & 20.12\% \\
annualised turnover (DEV window) & 14.1$\times$ & 15.1$\times$ \\
mean post-cap gross & 1.960 & 1.833 \\
Ulcer index & \textbf{0.0673} & 0.0698 \\
realized coverage / nominal & 0.7483 / 0.7500 & 0.7483 / 0.7500 \\
fraction of asset-day cells with a non-zero position & 0.980 & 0.980 \\
\bottomrule
\end{tabular}
\par\smallskip
\parbox{0.86\linewidth}{\footnotesize Turnover is the DEV-window mean of daily
$\sum_i|\Delta w_i|$, annualised. The development records carried
5.7$\times$/6.0$\times$, a full-sample average diluted by the low-activity
pre-2016 years; the window figures reconcile with the cost step of
Section~\ref{sec:robust} ($14.1\times5$\,bps $\approx0.71$\,pp). By block
bootstrap, Config A's growth has a standard error of 0.082 (95\% interval
$[0.136,0.454]$), against the analytic 0.095 quoted in
Section~\ref{sec:threat}; Config A minus Config B is $+2.6$\,pp
\ci{-0.6}{+6.4}.}
\end{table}

Config B's Ulcer index is \emph{worse} than Config A's despite its far smaller
maximum drawdown: the dial truncates the drawdown tail without reducing typical
drawdown.

\begin{table}[t]
\centering
\begin{minipage}[t]{0.47\linewidth}
\centering
\captionof{table}{Per-year net log growth, DEV.}
\label{tab:peryear}
\small
\begin{tabular}{@{}lcccc@{}}
\toprule
year & Config A & Config B & SPY bar & B7 \\
\midrule
2016 & +0.154 & +0.149 & +0.087 & +0.243 \\
2017 & +0.353 & +0.353 & +0.148 & +0.435 \\
2018 & $-$0.032 & $-$0.045 & $-$0.032 & $-$0.133 \\
2019 & +0.306 & +0.226 & +0.205 & +0.510 \\
2020 & +0.667 & +0.610 & +0.136 & +0.268 \\
2021 & +0.256 & +0.256 & +0.191 & +0.436 \\
\bottomrule
\end{tabular}
\end{minipage}\hfill
\begin{minipage}[t]{0.50\linewidth}
\centering
\captionof{table}{Per-asset decomposition, Config A, DEV: annualised
arithmetic net contribution, mean post-cap position, and per-asset coverage
(nominal 0.75).}
\label{tab:perasset}
\small
\begin{tabular}{@{}lccc@{}}
\toprule
asset & contribution & position & coverage \\
\midrule
SPY & +0.0585 & +0.322 & 0.726 \\
QQQ & +0.0836 & +0.334 & 0.690 \\
DIA & +0.0568 & +0.308 & 0.737 \\
MDY & +0.0383 & +0.265 & 0.729 \\
GLD & +0.0242 & +0.269 & 0.779 \\
SLV & +0.0021 & +0.114 & 0.781 \\
USO & +0.0406 & $-$0.039 & 0.781 \\
DBC & +0.0080 & $-$0.003 & 0.764 \\
\textbf{total} & \textbf{+0.3120} & +1.571 & 0.748 \\
\bottomrule
\end{tabular}
\end{minipage}
\end{table}

Development growth was positive in five of six years
(Table~\ref{tab:peryear}); 2018 is the loss year, shared with every comparator.
The second half is much stronger than the first (0.41 against 0.16 for Config
A), driven by 2020, the single largest concentration risk in the result.
Contributions in Table~\ref{tab:perasset} are arithmetic and sum to $+0.3120$;
the headline log growth 0.2845 is this less the variance drag
$\sigma^2/2=0.2336^2/2\approx0.0273$. All eight assets contribute positively
and the largest (QQQ) is 27\% of the total. USO earns $+0.0406$ from a
\emph{mean position of $-0.039$}, from timing rather than a static tilt, which
was the cleanest development-window evidence that the sizing rule does
something; on the sealed window USO was the one asset that lost
($-0.036$, Section~\ref{sec:lbdetail}). Per-asset coverage spans 0.690--0.781;
with 21-day overlapping targets the effective sample is $n\approx70$ per asset,
so one standard error is about 0.052 and QQQ at 0.690 is 1.2 standard errors
below nominal: not significant, but stated rather than buried.

\begin{figure}[t]
\centering
\includegraphics[width=0.88\linewidth]{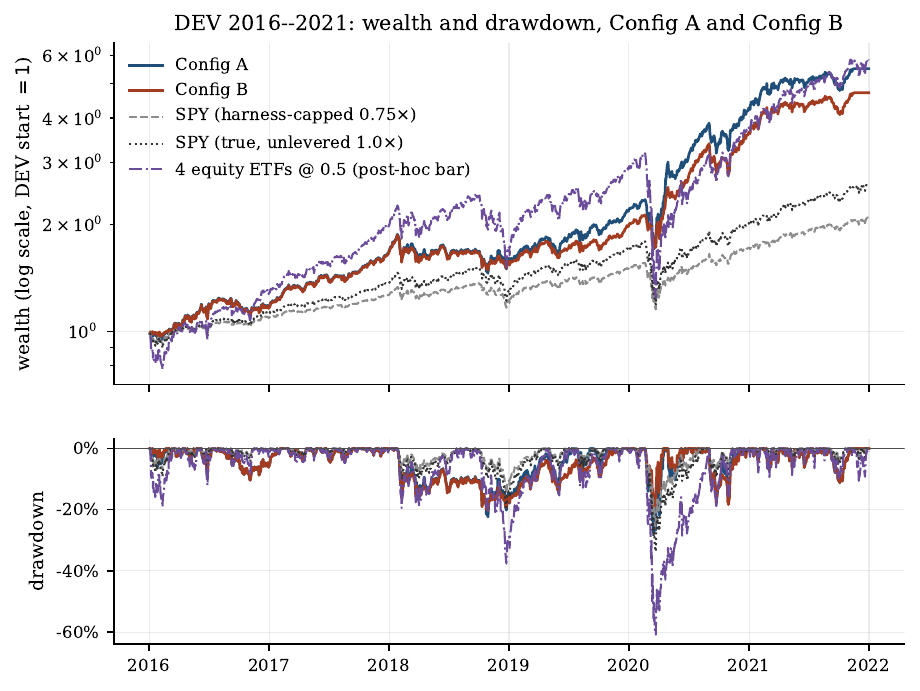}
\caption{DEV wealth (log scale) and drawdown for Config A and Config B, against
the harness-capped SPY bar ($0.75\times$), true unlevered SPY, and the post-hoc
four-equity book B7.}
\label{fig:devwealth}
\end{figure}

\subsection{What estimates $\sigma$}
\label{sec:sigma}

\begin{figure}[t]
\centering
\includegraphics[width=0.84\linewidth]{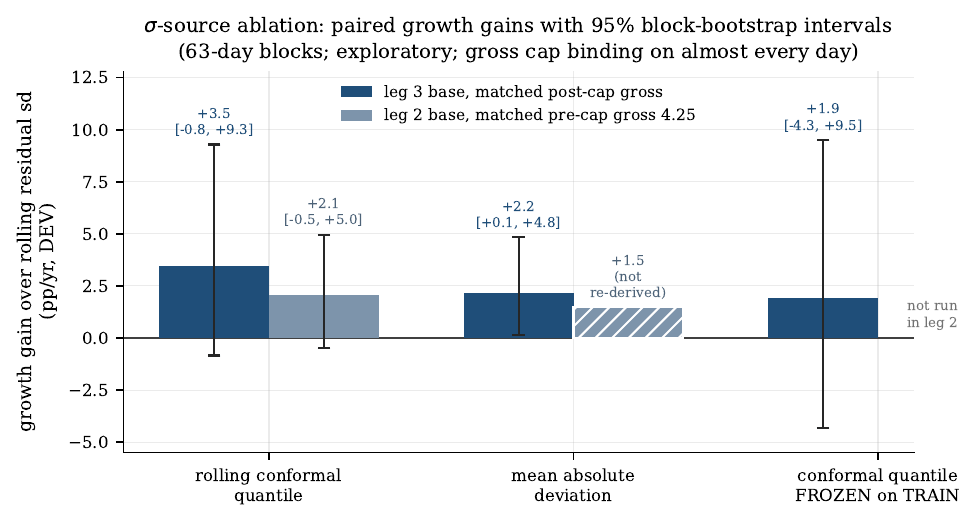}
\caption{The $\sigma$-source ablation as growth gains over the rolling residual
standard deviation, both ablation configurations as paired bars, with 95\%
moving-block-bootstrap intervals (63-day blocks).}
\label{fig:sigma}
\end{figure}

At matched leverage, window and causality, Table~\ref{tab:sigma} changes only
the functional of the calibration residuals in the Kelly denominator, and
Figure~\ref{fig:sigma} plots the gains with their intervals.

\begin{table}[t]
\centering
\caption{$\sigma$-source ablation, DEV. $\Delta$: growth difference against the
residual standard deviation of the same block, in pp, with a 95\%
block-bootstrap interval. Every row was measured with the 2.0 gross cap
binding almost always (post-cap gross 1.92--1.96); off the cap the ordering is
untested.}
\label{tab:sigma}
\small
\setlength{\tabcolsep}{4.5pt}
\begin{tabular}{@{}lccccl@{}}
\toprule
$\sigma$ estimator & growth & Sharpe & gross & source & $\Delta$ vs sd \\
\midrule
rolling conformal quantile $q_{0.75}(|e|)/z$ & \textbf{0.2821} & \textbf{1.34} & 1.960 & leg-3 base & $+3.5$ \ci{-0.8}{+9.3} \\
rolling mean absolute deviation & 0.2693 & 1.27 & 1.952 & leg-3 base & $+2.2$ \ci{+0.1}{+4.8} \\
conformal quantile \textbf{frozen on TRAIN} & 0.2669 & 1.21 & 1.961 & leg-3 base & $+1.9$ \ci{-4.3}{+9.5} \\
rolling residual standard deviation & 0.2476 & 1.19 & 1.922 & leg-3 base & --- \\
\midrule
rolling conformal quantile, matched pre-cap gross & \textbf{0.2774} & 1.32 & 1.959 & leg-2 base & $+2.1$ \ci{-0.5}{+5.0} \\
$1.2533\times$ mean absolute deviation & 0.2713 & 1.30 & 1.958 & leg-2 base & $+1.5$ (not re-derived) \\
residual standard deviation & 0.2566 & 1.24 & 1.957 & leg-2 base & --- \\
\bottomrule
\end{tabular}
\end{table}

Leverage matching is exact in the leg-2 block (post-cap gross 1.957--1.959) but
only approximate in leg 3, where the standard-deviation row runs 1.9\% lower
(1.922 against 1.960). A gross-matched replay (the Kelly fraction re-solved to
restore post-cap gross of about 1.958) scores the standard deviation at
0.2593--0.2609 and the mean absolute deviation at 0.2742: the ordering is
unchanged, but the leg-3 conformal-versus-sd gap narrows from 3.5 to roughly
2.2\,pp, and the exactly matched leg-2 figure of 2.1\,pp is the headline.

Two readings, both hedged by the intervals. First, in both ablation
configurations the growth ordering is exactly the estimators'
\textbf{tail-sensitivity} ordering: the standard deviation's influence function
is quadratic in the residual, the mean absolute deviation's is linear, and the
quantile's is bounded. A candidate mechanism, not measured directly: the
conformal quantile systematically \emph{understates} $\sigma$ for the
fat-tailed assets (USO, SLV), so the book keeps more weight on them, which pays
only because the gross cap holds the portfolio below its Kelly optimum. The
headline 2.1\,pp has a 95\% interval of $[-0.5,+5.0]$; of the contrasts in
Table~\ref{tab:sigma}, only mean absolute deviation over standard deviation
excludes zero.

Second: a $\sigma$ estimated \textbf{once} on pre-2016 residuals and held for
six years (0.2669) still beats a \emph{rolling} residual standard deviation
(0.2476) in point estimate, with a wide interval. Decomposing the 3.5\,pp gap,
the \textbf{updating} axis is worth $+1.5$\,pp \ci{-2.0}{+5.1} (rolling versus
frozen conformal); the remaining $+1.9$\,pp mixes the functional change with
the loss of updating and is not separately identified without a frozen-sd cell,
which we do not have. That is the honest bound on what the conformal machinery
is responsible for: on this window, the functional of the residuals mattered
more than updating it, and slower was never worse in point estimate.

\subsection{How much of the conformal cross-section to keep}
\label{sec:xsec}

Two dials bracket the answer from opposite sides; neither was re-derived, so
the values are point estimates.

\textbf{Flattening $\sigma$'s cross-section costs 6.6\,pp.} Replacing each
asset's $\sigma$ by the day's cross-sectional geometric mean at matched gross,
with $\theta$ interpolating (Table~\ref{tab:sweeps}a), loses monotonically.
The exponent in $f\sim\mu/\sigma^{\gamma}$ has an interior optimum
(Table~\ref{tab:sweeps}b): the sweep peaks at $\gamma\approx2$--$2.5$, a plateau
containing the textbook inverse-variance exponent; the argmax (2.5) is 0.1\,pp
above $\gamma=2$ and we read nothing into it.

\textbf{But the raw Kelly ratios must be winsorised.} At matched gross (clip at
$c$, renormalise to 2.0), the sweep is unimodal, with an interior optimum at
$c\approx0.75$--$0.90$ and a flat right tail once the clip ceases to bind
(Table~\ref{tab:sweeps}c). Discarding the Kelly magnitude entirely (sign-only,
0.2009 at Sharpe 1.02) and keeping it entirely (0.2413 at Sharpe 1.07) are
\emph{both} 4--8\,pp worse than winsorising it. \textbf{The harness's exogenous
\texttt{MAX\_PER\_ASSET = 0.75} happens to sit essentially at the optimum, which
is luck, not design.} Counter-intuitively, ``fixing'' the clip by
water-filling---renormalise first so unpinned assets keep their exact Kelly
ratios---is a 4.0\,pp disaster ($0.2821\to0.2417$, Sharpe $1.337\to1.066$,
maximum drawdown $0.281\to0.371$), while clip-then-renormalise is
metric-neutral (0.2808). On these point estimates the per-asset cap is not a
nuisance constraint here but a cross-sectional robustification.

\begin{table}[t]
\centering
\caption{Development-window sweeps at matched gross (point estimates). (a)
cross-sectional shrinkage of $\sigma$ toward the daily geometric mean
($\theta=1$ keeps the full cross-section); (b) the exponent $\gamma$ in
$f\sim\mu/\sigma^\gamma$; (c) the per-asset winsorisation level $c$.}
\label{tab:sweeps}
\small
\begin{tabular}{@{}l*{6}{c}@{}}
\toprule
(a) $\theta$ & 0 & 0.25 & 0.5 & 0.75 & 1 & \\
growth & 0.2113 & 0.2287 & 0.2527 & 0.2705 & 0.2773 & \\
\midrule
(b) $\gamma$ & 0.5 & 1 & 1.5 & 2 & 2.5 & 3 \\
growth & 0.2305 & 0.2518 & 0.2701 & 0.2773 & 0.2781 & 0.2722 \\
\midrule
(c) clip $c$ & 0.26 & 0.30 & 0.35 & 0.45 & 0.55 & 0.65 \\
growth & 0.2322 & 0.2394 & 0.2474 & 0.2601 & 0.2701 & 0.2772 \\
clip $c$ & \textbf{0.75} & \textbf{0.90} & 1.20 & 2.0 & 5.0 & 50 \\
growth & \textbf{0.2809} & \textbf{0.2819} & 0.2747 & 0.2477 & 0.2413 & 0.2413 \\
\bottomrule
\end{tabular}
\end{table}

\subsection{What each design choice contributed}
\label{sec:steps}

\begin{table}[t]
\centering
\caption{Design steps of the search, in the order adopted (point estimates).}
\label{tab:steps}
\small
\begin{tabular}{@{}lcL{6.6cm}@{}}
\toprule
change & $\Delta$ growth & note \\
\midrule
daily $\to$ 21-day forecast horizon & +7.5\,pp & $0.173\to0.248$; Sharpe $0.86\to1.22$; turnover $37.7\to17.9$ \\
position smoothing, EWMA(4) & +1.0\,pp & turnover $17.9\to9.2$ \\
calibration window $250\to500$ days & +0.4\,pp & \\
$\kappa$: $0.25\to0.15$ & +0.7\,pp & pure cap-saturation geometry \\
horizon ensemble $\{12,16,21,27,34\}$ & +0.7\,pp & broad plateau, not a spike \\
staggered books, $K=5$, replacing EWMA & +0.5\,pp & turnover $7.7\to7.0$ \\
conformal inverse-variance ensemble weights & $\approx0$ & turnover $6.96\to5.59$ \\
anchor shrinkage $\lambda=0.3$ & +0.24\,pp & max dd $0.2813\to0.2768$; coverage $0.7559\to0.7483$ \\
\bottomrule
\end{tabular}
\end{table}

The horizon change dominates by an order of magnitude (Table~\ref{tab:steps}):
at a one-day horizon the drift is buried in noise and the ridge produces a
near-market-neutral book that throws away the equity risk premium, while at a
monthly horizon the drift dominates and the bets become long-biased, with an
interior optimum at 21 ($h=42\to0.238$, $h=10\to0.222$). Everything after the
first two rows is small, and the last $\approx1$\,pp is documented tuning
expected to shrink out of sample, as the registration predicted.

\section{What does not work on the development window}
\label{sec:refute}

The search's second and third legs tested nine structural hypotheses and
rejected eight. The rejections share a sign and, we conjecture, a mechanism;
the intervals below say how much weight each can bear.

\subsection{Locally adaptive conformal prediction}
\label{sec:adaptive}

\begin{table}[t]
\centering
\caption{Devices tested and rejected on DEV. $\Delta$: growth against the base
configuration of the leg in which the device was tested, with a 95\%
block-bootstrap interval where the arm was re-derived (pp/yr).}
\label{tab:adaptive}
\small
\begin{tabular}{@{}L{5.6cm}L{5.0cm}L{3.6cm}@{}}
\toprule
device & result & reference \\
\midrule
volatility-scaled nonconformity score & $-5.3$\,pp at $h=1$; $-3.3$\,pp at $h=21$ & standard in financial conformal work \\
adaptive conformal inference, step $\gamma\in[0.001,0.05]$ & monotonically worse; $-1.6$ \ci{-5.1}{+1.6} at $\gamma=0.005$; 0.128 at $\gamma=0.05$ & \citet{gibbs2021} \\
recency-weighted conformal, decay $\rho\in[0.98,0.999]$ & monotone toward $\rho=1$; $-1.4$ \ci{-4.3}{+1.5} at $\rho=0.99$ & \citet{barber2023} \\
asymmetric two-sided intervals with a downside-only $\sigma$ & $-1.6$ \ci{-3.6}{+0.2} & CQR-flavoured \\
pooled calibration across all eight assets (leg-1 construction) & $-4.2$ & \\
pooled calibration across all eight assets (leg-2 code path) & $-2.5$ \ci{-6.6}{+0.7}; 0.2573 & \\
Mondrian calibration within asset class & $-2.4$ \ci{-6.8}{+1.1}; 0.2584 & \citet{vovk2003} \\
sizing on the $H_s$-day holding-period residual & monotone in $H_s$; $-0.7$ \ci{-2.0}{+0.5} at $H_s=5$; $-1.5$ at $H_s=1$ & \\
\bottomrule
\end{tabular}
\end{table}

Every one of the first four devices in Table~\ref{tab:adaptive} makes the
interval adapt faster, and every one loses growth in point estimate; none of
those losses is individually distinguishable from zero. Our candidate
explanation is that the sizing map integrates the interval and so charges the
\emph{variance} of $q$ rather than rewarding its local sharpness. We measured
that mechanism directly only for the holding-period device, which calibrates
and sizes on the $H_s$-day residual the staggered book is actually exposed to,
instead of the 21-day forecast residual. There the estimator's smoothness is
analytically transparent: an overlapping 21-day residual is itself a 21-day
moving average of daily shocks, so its trailing quantile is smoother, with the
standard deviation of daily log-changes in $q$ at 0.00343 for $H_s=21$ against
0.00391 at $H_s=5$ and lag-1 autocorrelation 0.53 against 0.36, and growth is
monotone in $H_s$. For the other devices the mechanism is conjectured, not
measured. The pooled rows fail for a different reason, taken up below.

One correction to the development notes: what fails is \emph{fast}
adaptivity, not cross-sectional adaptivity---Section~\ref{sec:xsec} shows that
flattening the cross-section of $\sigma$ costs 6.6\,pp. On this evidence it is
the speed of the scale estimator, not its cross-sectional content, that should
be slow. Mondrian calibration sharpens this: pooling within asset class
(0.2584) recovers only 0.1\,pp over all-asset pooling through the identical code
path (0.2573), so the original diagnosis (``score shapes differ across
assets'') was wrong. The $-4.2$\,pp figure is a different leg-1 construction
and is not the right comparator. Pooling fails, we think, because it replaces
each asset's own quantile with (its trailing mean absolute deviation)
$\times$ (a shape factor shared across the cell), discarding the per-asset tail
behaviour that Section~\ref{sec:sigma} suggests matters.

\subsection{Holding-period structure}
\label{sec:holding}

Staggered overlapping books beat every equal-memory EWMA ($K=3/4/5\to
0.2809/0.2814/0.2820$ against EWMA span $4/8/13/21\to
0.2774/0.2714/0.2667/0.2617$), so it is the flat kernel, not the length, that
helps. But \textbf{long staggering fails badly} ($K=21\to0.2545$,
$K=63\to0.2383$): the 21-day forecast decays with a roughly one-week half-life,
so ``hold the book for the forecast horizon'' is wrong here. The mismatch is
real and is \emph{not} fixed by matching the calibrated object to the traded
one (the last row of Table~\ref{tab:adaptive}, which loses).

\subsection{The conformal predictive decision rule}
\label{sec:cps}

Solving $f^\star=\arg\max_f\mathbb{E}[\log(1+fR)]$ over the empirical
calibration-residual distribution $\{\hat\mu_t+e_i\}$---Vovk and Bendtsen's
expected-utility maximisation, which under log utility \emph{is}
Kelly---loses in every variant we could construct. What we implement is that
decision rule rather than a full conformal predictive system, so the
comparison is against the practical version of the idea, not its most refined
form. Raw, it scores \textbf{0.1525 at Sharpe 0.74}; the best variant
(mean-recentred) reaches 0.2411, still $-3.6$\,pp \ci{-6.7}{-0.4} against its
base of 0.2774, the one device whose loss excludes zero. Two measured problems:
\begin{enumerate}
\item \textbf{The predictive distribution's location is dominated by forecast
bias.} The trailing 500-day mean residual is the \emph{same order as the point
forecast itself} (SPY: mean $|\hat\mu|=0.0071$ against $|$mean
residual$|=0.0048$; USO: 0.0192 against 0.0216). Recentring swamps everything
else: raw 0.1525, mean-recentred 0.2411, median-recentred 0.0581 at a 59\%
maximum drawdown.
\item \textbf{Even recentred, empirical-predictive-distribution Kelly charges
the realized left tail}, which shrinks fat-tailed assets exactly as
inverse-volatility tilting does (the first row of Table~\ref{tab:adaptive},
reached by another route).
\end{enumerate}
A companion result sharpens the point: conformal probability-of-profit sizing,
$f=(2p-1)/\sigma$ with $p=P_{\text{CPD}}(R>0)$, the most tail-robust member of
the family, scores 0.234--0.246 with maximum drawdown blowing out to
0.40--0.53. Robustness appears to help in the denominator and hurt in the
numerator: passing the forecast through the calibration distribution destroys
its \emph{magnitude}, and Kelly needs magnitude.

\subsection{Correlation}
\label{sec:corr}

Conformalising the realized \emph{portfolio} residual and solving one Kelly
leverage for the whole book merely traces the leverage curve and converges back
to the per-asset answer once the cap binds (portfolio $\kappa=0.05\to1.0$:
0.098/0.170/0.228/0.271/0.276/0.264/0.249). Multivariate Kelly
$w=\Sigma^{-1}\mu$ is far worse (0.023--0.179 across shrinkage 0--0.9) at
turnover 11--20$\times$: Markowitz instability plus a hedged book that discards
the equity premium. \textbf{Correlation is the biggest theoretical hole in the
per-asset formulation and is empirically inert here}: under a binding gross cap
only the \emph{direction} survives. That is a limitation of the study design,
not evidence about the method off the cap.

\subsection{Two apparent gains that did not survive their own controls}
\label{sec:honesty}

\textbf{Shrinkage drift estimators.} Grand-mean pooling of the drift at weight
0.10--0.30 appears to gain (0.2854--0.2874, a broad plateau), but the control
kills it: a \textbf{fixed constant} in place of the data-driven grand mean does
as well or better (0.2854--0.2904 for $c=0.002\dots0.020$). It is a hand-tuned
long tilt with no data in it. \textbf{Not adopted}, and reported, because it is
exactly the kind of finding that would not survive a sealed window and usually
goes unreported.

\textbf{Epistemic model disagreement.} Adding $w\cdot\mathrm{disp}^2$ to the
denominator (disp is the dispersion of the five horizon components) is
monotone in $w$ with no interior optimum, and its control---disp replaced by its
own causal expanding per-asset mean, same cross-sectional tilt, zero
timing---does just as well. The diagnostic: \textbf{mean ensemble disagreement
is 0.6--2.5\% of the mean conformal half-width} (SPY 0.00053 against 0.0365;
USO 0.00297 against 0.1187), so epistemic uncertainty is about 1.5\% of
predictive uncertainty and the Kelly denominator is essentially purely
aleatoric. \textbf{Not adopted}, despite $+0.1$\,pp.

\section{The drawdown dial and its two controls}
\label{sec:dial}

Trailing conformal miscoverage is an empirical regime indicator---a ``the
predictive distribution has shifted'' signal calibrated against its own history
rather than by any distribution-free guarantee (Section~\ref{sec:method})---and
the leading hypothesis was that de-levering on it should raise growth.
\textbf{It does not.} Across 40+ configurations spanning window 21--252 and
sensitivity 0.5--2, every variant loses growth, monotonically in how much gross
it cuts, and the leverage-matched version ($\kappa$ re-solved to restore
post-cap gross 1.960) still loses 0.8--2.5\,pp. Not a leverage tax, then: the
\emph{timing} is bad, and measurably so---trailing miscoverage peaks
\emph{after} a shock, during the rebound, so it sells the bottom (2018 Q4,
2020 Q1).

The same statistic reduced drawdown. Table~\ref{tab:beta} gives the frontier
over $\beta$ at $M=21$ on Config A; coverage is 0.7483 at every point, since
the dial never touches the interval.

\begin{table}[t]
\centering
\caption{The drawdown dial on DEV: frontier over $\beta$ at window $M=21$.
Turnover is the full-sample average carried by the development records
(Section~\ref{sec:dev}).}
\label{tab:beta}
\small
\begin{tabular}{@{}lccccccc@{}}
\toprule
$\beta$ & growth & Sharpe & max dd & Ulcer & Calmar & turnover & post-cap gross \\
\midrule
0 (Config A) & 0.2845 & 1.336 & 0.2768 & 0.0673 & 1.127 & 5.69 & 1.960 \\
0.25 & 0.2595 & 1.311 & 0.2583 & 0.0710 & 1.096 & 5.98 & 1.915 \\
0.50 & 0.2551 & 1.335 & 0.2313 & 0.0768 & 1.196 & 6.06 & 1.872 \\
0.75 & 0.2503 & 1.337 & 0.2207 & 0.0718 & 1.228 & 6.07 & 1.843 \\
\textbf{1.00 (Config B)} & \textbf{0.2584} & \textbf{1.386} & \textbf{0.2026} & 0.0698 & \textbf{1.376} & 6.03 & 1.833 \\
1.25 & 0.2576 & 1.386 & 0.2034 & 0.0700 & 1.366 & 6.17 & 1.823 \\
1.50 & 0.2563 & 1.382 & 0.1991 & 0.0687 & 1.388 & 6.26 & 1.816 \\
2.00 & 0.2556 & 1.385 & 0.1994 & 0.0693 & 1.382 & 6.15 & 1.802 \\
3.00 & 0.2586 & 1.411 & 0.1896 & 0.0633 & 1.467 & 6.27 & 1.787 \\
\bottomrule
\end{tabular}
\end{table}

\begin{figure}[t]
\centering
\includegraphics[width=0.92\linewidth]{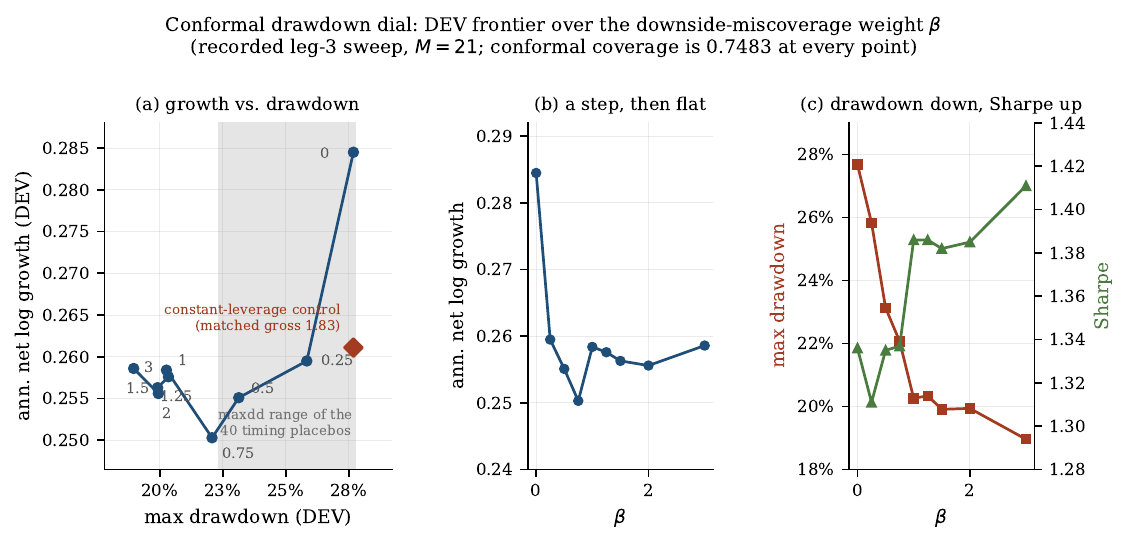}
\caption{Growth--drawdown frontier over the downside-miscoverage weight
$\beta$ on DEV, with the constant-leverage control and the placebo range.}
\label{fig:beta}
\end{figure}

\textbf{This is a step, not a smooth efficient frontier.} Growth drops about
2.5\,pp the moment the dial is switched on at all and is then flat in $\beta$
(0.250--0.259 across $\beta\in[0.25,3]$), while maximum drawdown falls from
27.7\% to 19.0\%, monotonically to within the noise of the last three grid
points, and Sharpe \emph{rises} from 1.34 to 1.41: a fixed entry toll, then
drawdown almost for free (Figure~\ref{fig:beta}). $\beta=1$ is reported as the
parameter-free point, and it dominates $\beta=0.5$ on every axis. The window
$M$ is second-order and noisy (at $\beta=1$, $M=10/15/21/31/42\to$ growth
0.249/0.257/0.258/0.252/0.254, maximum drawdown
0.230/0.198/0.203/0.240/0.233). By block bootstrap, the dial's growth cost on
DEV, Config A minus Config B, is $+2.6$\,pp \ci{-0.6}{+6.4}.

\textbf{Which statistic matters}, at $M=21$, $\beta=1$: one-sided
\emph{downside} miscoverage 0.2584 / max dd 0.203; net-directional 0.2573 /
0.201; two-sided pooled 0.2355 / 0.236; interval \emph{width} 0.2785 / 0.2768
(barely acts). Only the signed, one-sided break rate carried usable risk
information on DEV.

\textbf{Control 1: constant leverage at matched gross.} Reaching the dial's
realized gross of 1.83 requires a constant multiplier of 0.45, giving growth
0.2611, Sharpe 1.273 and \textbf{maximum drawdown 0.2768, completely
unchanged}: the gross cap absorbs any uniform shrink until the cut is very
deep. At equal growth and equal realized leverage the conformal dial buys
$-7.4$\,pp of drawdown and $+0.11$ of Sharpe, so on DEV the timing, not the
average de-levering, is doing the work.

\textbf{Control 2: circular-shift placebo.} Circularly shifting the dial's own
multiplier series 40 ways preserves the marginal distribution of cuts and
destroys the timing. The realized maximum drawdown of 0.2026 is beaten by
\textbf{0 of 40} (null median 0.2768, minimum 0.2234), rank-based
$p=(0+1)/(40+1)=0.024$; Sharpe is beaten by 3 of 40, $p=4/41=0.098$
\citep{phipson2010}. Because the dial, its window and this statistic were
selected on DEV before this test was run, and the circular-shift null assumes
shift-invariance of the sequence, these are exploratory placebo ranks rather
than confirmatory significance. Two results cut the other way and are
reported: the realized growth of 0.2584 is beaten by \textbf{31 of the 40
shifts} (rank-based $p=0.78$), so the timing \emph{costs} growth relative to
random cutting of the same size, and the \textbf{Ulcer index does not improve}
(0.0698 against a placebo mean of 0.0622). The dial truncates the drawdown tail;
it does not reduce typical drawdown and it does not improve growth. On the
sealed window it failed its registered criterion
(Section~\ref{sec:interpret}).

A structural corollary: the identical statistic applied \textbf{per asset}
rather than pooled scores 0.2361 at maximum drawdown 0.317 and turnover 9.2,
plausibly because a rolling mean of $M$ binary indicators (about 2.6 expected
breaks in 21 days) is nearly pure noise. The regime statistic must be pooled across
assets while the scale statistic stays per asset: the two conformal objects
want opposite aggregation.

\section{Robustness and the honesty ledger}
\label{sec:robust}

\textbf{Cost sensitivity.} At 0, 5, 10, 20 and 50\,bps per unit of turnover,
Config A scores 0.2916, 0.2845, 0.2775, 0.2634 and 0.2212 (Sharpe 1.366, 1.336,
1.305, 1.245 and 1.064) and Config B 0.2660, 0.2584, 0.2508, 0.2357 and
0.1904.
On the development window the result is not a cost-model artifact: it survives
a $10\times$ cost increase on eight of the most liquid ETFs listed. The window
turnover behind these numbers is 14.1$\times$/yr (Config A) and 15.1$\times$/yr
(Config B), which reconciles with the 0.71\,pp step above; the 5.7$\times$
carried by the development records is a full-sample average diluted by the
low-activity pre-2016 years. Sealed-window turnover was 16.0$\times$, and the
$10\times$-cost claim does not carry over there (Section~\ref{sec:lbdetail}).

\textbf{Financing.} The harness charges nothing for the roughly 0.98 units of
borrowed exposure carried daily. Charging 1/2/4\% a year on gross above 1.0
costs Config A 0.98/1.96/3.91\,pp of growth ($0.2845\to0.2747/0.2650/0.2454$)
and Config B 0.85/1.70/3.41\,pp; the comparison against \emph{leveraged}
passive books is untouched, since those books borrow the same amount
(\texttt{paper/financing\_overlay.md}).

\textbf{Placebo lag.} An \emph{extra} implementation lag decays the metric
gracefully---$+1$\,d 0.2793, $+2$\,d 0.2729, $+3$\,d 0.2637, $+5$\,d 0.2381,
$+10$\,d 0.2279---rather than collapsing: the signature of a slow signal, not
a look-ahead bug.

\textbf{$\alpha$ is flat once leverage is controlled.} Re-solving $\kappa$ at
each $\alpha$ to hold pre-cap gross at 4.25, Sharpe is 1.30--1.33 throughout
and only $\alpha\le0.10$ degrades growth materially (Table~\ref{tab:alpha}).
The shallow maximum at $\alpha=0.35$ ($+0.3$\,pp) was \textbf{not} adopted: it
buys noise at the cost of a much weaker nominal level.

\begin{table}[t]
\centering
\caption{Gross-matched $\alpha$ sweep on DEV (point estimates).}
\label{tab:alpha}
\small
\begin{tabular}{@{}l*{9}{c}@{}}
\toprule
$\alpha$ & 0.05 & 0.10 & 0.15 & 0.20 & 0.25 & 0.30 & 0.35 & 0.40 & 0.45 \\
growth & .2676 & .2727 & .2735 & .2756 & .2773 & .2775 & .2804 & .2798 & .2767 \\
\bottomrule
\end{tabular}
\end{table}

\textbf{Exchangeability.} The 21-day overlap makes consecutive nonconformity
scores strongly dependent, the standard objection to the guarantee here.
Restricting calibration to non-overlapping residuals (stride $=H=21$,
$n\approx23$) costs 1.3\,pp but leaves realized coverage essentially unchanged
(0.738 against 0.756 at nominal 0.75). The overlap therefore does not visibly
break \emph{marginal} coverage on DEV, and the overlapping estimator wins,
plausibly because its variance reduction beats its effective-sample-size loss. We do not
claim the finite-sample guarantee holds; Section~\ref{sec:split} shows how
little marginal coverage says about coverage in a given regime.

\textbf{Causality.} Re-verified line by line after each restructure: the forward
target is \texttt{rolling(h).sum()\allowbreak.shift(-h)} (an arithmetic sum of daily returns
rather than a compounded return, a second-order difference identical across
every compared configuration); training truncated to labels landed by the
prediction row; conformal window $[t-h-W+1,\,t-h+1)$; the expanding anchor at
row $t$ reads \texttt{sr[:t-h+1]}, recomputed every 21 rows and held stale;
regime indicators shifted by $H$ to their landing dates. During the
development legs no data after 2021-12-31 was ever loaded and
\texttt{FINAL\_EVAL} was never set; the single unsealing reported in
Section~\ref{sec:lockbox} happened only after the loop was closed and the
registration committed.

\begin{figure}[t]
\centering
\includegraphics[width=0.86\linewidth]{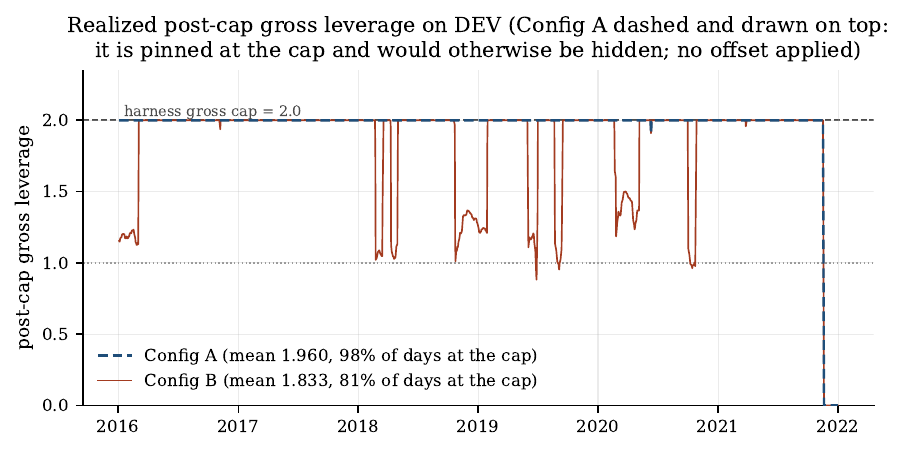}
\caption{Realized post-cap gross leverage on DEV; Config A dashed and drawn on
top, pinned at the cap on 98\% of days. Under the cap the interval sets the
book's composition, not its size.}
\label{fig:gross}
\end{figure}

\textbf{A known artifact, disclosed.} Both frozen configurations carry
\texttt{predict\_tail=False}, which suppresses a forecast wherever a horizon
component's label has not yet landed and so blanks the end of the processed
sample---visible as gross leverage dropping to zero in December 2021 in
Figure~\ref{fig:gross}. The causally correct \texttt{predict\_tail=True} scores
0.2789 at coverage 0.7487, with 98.9\% of asset-day cells carrying a non-zero
position, under the DEV truncation: it \emph{costs} 0.6\,pp there because late
December 2021 was adverse. With the truncation removed the two variants are
identical on DEV (0.2813; Section~\ref{sec:lockbox}) and differ only at the end
of the sample, where the secondary trades 13 more days. The registration
commits to reporting both variants on the sealed window.

\subsection{The open threat: a post-hoc four-equity book beats us on
development growth}
\label{sec:threat}

Four equity ETFs held at 0.5 each---gross 2.0, no forecasting, no conformal
prediction, no rebalancing beyond drift---compounds at \textbf{0.2934 on DEV,
beating Config A's 0.2845.} We state that without qualification.

\emph{The book's case.} It requires nothing: no model, no calibration, no
forecast, no dial. If the apparatus of Section~\ref{sec:method} cannot beat four
index funds at $2\times$, the natural reading is that it is a complicated way of
being long equities under a leverage cap, and the decomposition partly
supports that. The four equity ETFs supply $+0.237$ of Config A's $+0.312$ total
arithmetic contribution, and Config A's mean equity position is $+1.23$ of its
$+1.57$ net exposure. \textbf{This strategy is substantially a long-equity
book.}

\emph{Why it is not a fair comparator.} First, it is \textbf{post hoc}: which
four of the eight assets to hold was chosen after seeing DEV. The identical rule
on the other block---four commodity ETFs at 0.5---scores \textbf{0.0892 at a
68.7\% drawdown}, and the strongest hindsight equity book available under the
caps (QQQ at the cap, the rest sharing the remainder) reaches \textbf{0.3119}.
The book is not a strategy but a readout of what hindsight asset selection is
worth here, at least 20\,pp of annual growth. Second, on DEV \textbf{its risk is
in a different regime}: maximum drawdown \textbf{60.9\%} against 27.7\% and
20.3\%; Sharpe 0.963 against 1.336 and 1.386; Calmar 0.605 against 1.127 and
1.376; volatility 38.2\% against 23.4\% and 20.1\%; worst year $-0.133$ against
$-0.032$. A 61\% drawdown at $2\times$ on borrowed money would in practice mean
margin calls and forced deleveraging well before the trough, so the book's paper
growth assumes a survivability it does not demonstrate. (The development notes
estimated this book's drawdown at ``$\sim$35\%''; the recomputed figure is
60.9\%. The estimate was wrong in our favour and we correct it.)

\emph{What we do not claim.} Not that we beat the post-hoc book on DEV growth,
and not that 0.2845 against 0.2934 is meaningful in either direction: over six
years at 23\% volatility one standard error on annualised growth is about 0.095
analytically (0.082 by block bootstrap), and the paired difference is
$-0.9$\,pp \ci{-17.7}{+22.2}. Version~1 argued that the risk-adjusted separation
was large, was not a post-hoc selection, and was the comparison that matters
when the objective is compounded wealth under a real leverage constraint. The
registration committed us to reporting this book on the sealed window
regardless (B7), and there the risk-adjusted defence failed
(Section~\ref{sec:interpret}).

\section{The agentic search process, disclosed}
\label{sec:agentic}

The configuration space was searched by an autonomous LLM agent across three
relay sessions (``legs'') against an immutable harness. Approximately
\textbf{200 configurations were scored on DEV}; \texttt{results.tsv} records 47
rows---\textbf{33 keep/discard experiment decisions} (18 keeps, 15 discards), 2
paper-configuration rows, 10 ablation blocks and 2 robustness blocks---with the
rest scratch sweeps inside an experiment. Roughly 25 structurally distinct
hypotheses were tested; legs 2 and 3 rejected eight of nine between them. The
same workflow extended to this manuscript: the text was drafted with LLM
assistance, then reviewed, verified line by line against the committed
experimental record, and edited by the author, who takes sole responsibility
for every claim in it.

We disclose this for three reasons, and it is a strength rather than a
confession only because of the third.

\textbf{First, it is a multiple-comparisons exposure.} Two hundred
configurations on 1{,}511 days of a single universe is a large search. The
dominant gain ($+7.5$\,pp from the forecast horizon) is mechanistic; the final
$\approx1$\,pp is tuning and should be expected to shrink out of sample
(Section~\ref{sec:steps}), as the registration predicted. The bootstrap
intervals of Appendix~\ref{app:postreg} are not adjusted for this search.

\textbf{Second, the search produced a \emph{kind} of result a smaller one would
not.} The value is not the $+11$\,pp of accumulated metric but the rejected
structural hypotheses, each with a controlled ablation. Testing conformal
predictive systems, portfolio-level conformalisation, Mondrian calibration,
holding-period residuals, per-asset regime dials, epistemic disagreement terms,
drift shrinkage and water-filled clipping, each with its own control, is
roughly a person-month of work. The negative results are a large part of the
paper.

\textbf{Third, and only because of this: the sealed window.} An agentic search
is only as credible as the evaluation it cannot see. Before unsealing,
\texttt{paper/prereg.md} fixed the two commit SHAs, the metrics, the nine
comparison books including the adversarial one, the sample-size caveat, the
mapping from outcomes to conclusions, a written prediction that Config A would
land materially below 0.2845, and a commitment to report both configurations
and both \texttt{predict\_tail} variants. A large disclosed search plus a small
sealed pre-registered test is a stronger epistemic position than an
undisclosed human search of the same size, which is the field's status quo.

A limitation of the registration itself: the project repository is private, so
the registration and its ordering relative to the unsealing are verifiable
through the committed history, which we can provide on request, rather than
through a public archive or timestamp. Readers should weight the registration
accordingly.

\section{Pre-registration and the sealed-window result}
\label{sec:lockbox}

\subsection{The registration}
\label{sec:prereg}

Before unsealing, \texttt{paper/prereg.md} fixed:
(1) the two configurations by git tag and commit SHA, neither modifiable;
(2) the one permitted code change---removing the research-time
\texttt{DEV\_END} truncation so the walk-forward can proceed---with the runner
printing an auditable \texttt{git diff} against each tag;
(3) ten metrics, plus six registered diagnostics (per-year growth, per-asset
contribution and coverage, the monthly distribution, cost sensitivity and the
financing overlay);
(4) the nine comparison books B1--B9 of Table~\ref{tab:l2}, including B7, the
post-hoc four-equity book that beat Config A on DEV, registered so that the
paper could neither quietly drop it nor claim credit for a narrow win, and B8
and B9, the same kind of hindsight object;
(5) a sample-size caveat: at 23\% volatility over about 2.7 years one standard
error on annualised growth is roughly $0.234/\sqrt{2.72}\approx0.14$, so the
window cannot distinguish a growth rate of 0.28 from one of 0.14, or from one
of 0.42;
(6) the mapping from outcomes to conclusions, under which coverage within
$\pm0.05$ of nominal and Config A growth positive and above the unlevered books
B1--B3 would confirm the central claim, coverage holding while Config A
underperformed the leveraged books B4--B6 would be a partial refutation, Config
B's drawdown strictly below Config A's at Sharpe no worse would confirm the
dial, and a B7 win would be reported in the abstract; and
(7) a written prediction that Config A's growth would land materially below its
DEV value while coverage transferred.

Seven commitments bound the aftermath: \textbf{one run}, executed once and
committed verbatim; \textbf{no re-tuning}---no \texttt{CFG} change, no new
configuration scored, no variant introduced and no new experiment on any window
after unsealing; \textbf{both configurations reported unconditionally}, in the
abstract and the results table; \textbf{all nine books reported}; the DEV
sections frozen before unsealing; negative results published; and every number
traceable to the single run's output. The registration explicitly did not
permit a second universe or snapshot, re-estimation on sealed data, reporting
only the better configuration or variant, adding a comparison book after the
fact, \textbf{splitting the window and reporting a favourable sub-period}, or
any sanity check against sealed data before the official run.

One amendment was registered before unsealing
(\texttt{paper/prereg\_amendment.md}): the \texttt{DEV\_END} truncation
interacts with \texttt{predict\_tail=False}, so the full-sample artifacts the
sealed run scores correspond to DEV growth 0.2813 (A) and 0.2552 (B) rather
than the tagged 0.2845 and 0.2584, and the primary and secondary variants are
identical on DEV. The registered protocol was executed once, on 2026-07-30
(\texttt{paper/lockbox\_run.py}; verbatim output committed as
\texttt{paper/lockbox\_results.txt}, with the auditable one-line diffs against
both tags printed in it).

\subsection{Results}
\label{sec:lbresults}

\begin{figure}[t]
\centering
\includegraphics[width=0.86\linewidth]{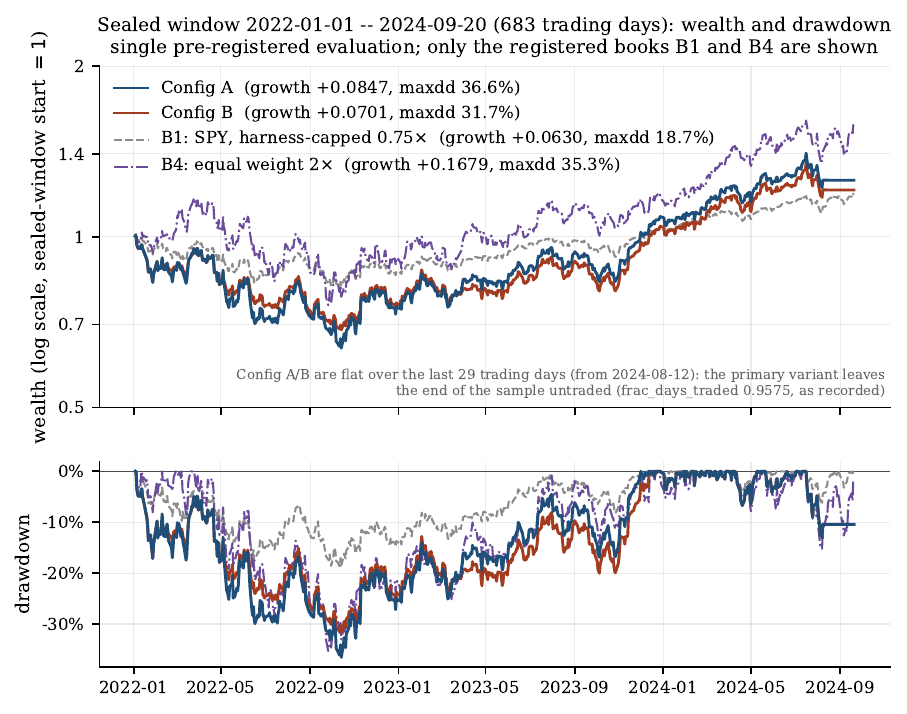}
\caption{Sealed-window wealth (log scale) and drawdown, Config A and Config B
(primary variant), with the registered books B1 (SPY, harness-capped
$0.75\times$) and B4 (equal weight $2\times$).}
\label{fig:lbwealth}
\end{figure}

Table~\ref{tab:l1} reports both configurations and both pre-declared variants,
Table~\ref{tab:l2} the nine registered books from the same run, and
Figure~\ref{fig:lbwealth} the wealth paths of both configurations against the
registered books B1 and B4.

\begin{table}[t]
\centering
\caption{Sealed window, 2022-01-01 to 2024-09-20 (683 trading days): both
configurations and both pre-declared variants, as registered.}
\label{tab:l1}
\small
\begin{tabular}{@{}lcccc@{}}
\toprule
 & A primary & B primary & A secondary & B secondary \\
\midrule
annualised net log growth & \textbf{+0.0847} & +0.0701 & +0.1147 & +0.0968 \\
Sharpe & 0.453 & 0.422 & 0.565 & 0.537 \\
maximum drawdown & 36.6\% & 31.7\% & 36.6\% & 31.7\% \\
Calmar & 0.327 & 0.303 & 0.410 & 0.389 \\
annualised volatility & 26.4\% & 22.8\% & 26.6\% & 22.9\% \\
annualised turnover & 16.0$\times$ & 17.6$\times$ & 16.2$\times$ & 17.8$\times$ \\
mean gross leverage & 1.91 & 1.72 & 1.95 & 1.75 \\
\textbf{realized coverage} / nominal 0.75 & \textbf{0.7450} & \textbf{0.7450} & \textbf{0.7485} & \textbf{0.7485} \\
$n$ coverage cells & 5{,}192 & 5{,}192 & 5{,}296 & 5{,}296 \\
fraction of asset-day cells with a non-zero position & 0.958 & 0.958 & 0.977 & 0.977 \\
Ulcer index & 0.162 & 0.159 & 0.161 & 0.158 \\
\bottomrule
\end{tabular}
\end{table}

\begin{table}[t]
\centering
\caption{The nine registered comparison books on the sealed window, scored in
the same single run.}
\label{tab:l2}
\small
\begin{tabular}{@{}llcccccc@{}}
\toprule
 & book & growth & Sharpe & max dd & Calmar & ann.\ vol & gross \\
\midrule
B1 & SPY buy \& hold (harness-capped at 0.75) & +0.0630 & 0.54 & 18.7\% & 0.39 & 13.5\% & 0.75 \\
B2 & equal weight, gross 1 & +0.0950 & 0.71 & 18.8\% & 0.56 & 14.9\% & 1.00 \\
B3 & vol-target risk parity 10\% & +0.0865 & 0.90 & 12.3\% & 0.74 & 10.2\% & 0.83 \\
B4 & equal weight $2\times$ & +0.1679 & 0.71 & 35.3\% & 0.60 & 29.7\% & 2.00 \\
B5 & inverse-vol risk parity $2\times$ & +0.1576 & 0.71 & 35.3\% & 0.55 & 27.3\% & 2.00 \\
B6 & vol-target risk parity 30\% & +0.1557 & 0.75 & 32.6\% & 0.57 & 24.9\% & 1.90 \\
B7 & \textbf{4 equity ETFs @ 0.5 (the threat)} & \textbf{+0.1079} & 0.48 & 46.2\% & 0.38 & 37.0\% & 2.00 \\
B8 & 4 commodity ETFs @ 0.5 (control) & \textbf{+0.1723} & 0.64 & 42.8\% & 0.58 & 38.7\% & 2.00 \\
B9 & equity, QQQ at the per-asset cap & +0.1082 & 0.47 & 48.4\% & 0.38 & 38.6\% & 2.00 \\
\bottomrule
\end{tabular}
\end{table}

\subsection{Interpretation, per the registered mapping}
\label{sec:interpret}

\begin{enumerate}[leftmargin=*]
\item \textbf{Coverage is consistent with nominal: the calibration half of the
registered criterion.} Coverage of 0.7450 (primary) and 0.7485 (secondary)
deviates from nominal by 0.005. Version~1 estimated, by a rough effective-sample
argument, a standard error of 0.03--0.05; a moving-block bootstrap gives 0.042
(63-day blocks; 0.029 and 0.052 at 21 and 126 days). The registered $\pm0.05$
band is therefore about one standard error wide: coverage is consistent with
nominal, not established to high precision. And coverage transferred
\emph{pooled over the window}; Section~\ref{sec:split} shows that it did not
transfer through 2022 itself.

\item \textbf{The growth claim takes the pre-declared partial refutation.}
Config A primary ($+0.0847$) is positive and beats B1 ($+0.0630$) but neither B2
($+0.0950$) nor B3 ($+0.0865$), and sits well below the naive leveraged books
B4--B6 ($+0.156$ to $+0.168$). The registered confirmation criterion for the
central claim is therefore not met. In the registered language: \emph{the
conformal interval is well calibrated but the sizing map does not beat naive
leverage out of sample} on this window. The secondary variant ($+0.1147$) beats
B1--B3 but remains below B4--B6; per the registration it cannot be promoted to
primary. The 3.0\,pp gap between the variants is mechanical: the two books are
identical except at the end of the sample, where the primary's
label-availability rule leaves no position held over the last 29 trading days
(from 12 August 2024) and the secondary over the last 16, a difference of 13
trading days, recorded as a
fraction of asset-day cells with a non-zero position of 0.958 against 0.977.
Those days were positive for the book, and annualising over a 2.7-year window
amplifies them; the swing sits well inside the registered one-standard-error
band of $\pm0.14$ ($\pm0.16$ at the realized volatility of 26.4\%) and should
be read as sampling noise around one result, not as two strategies. Paired
against the books, the window cannot settle signs: Config A minus B4 is
$-8.3$\,pp \ci{-30.0}{+10.8}, minus B2 $-1.0$\,pp \ci{-22.9}{+19.0}, minus B7
$-2.3$\,pp \ci{-16.4}{+12.0} (Appendix~\ref{app:postreg}).

\item \textbf{The drawdown dial is not confirmed under its strict criterion.} It
reduced maximum drawdown (31.7\% against 36.6\%) and the Ulcer index, but
Sharpe fell (0.42 against 0.45), failing the pre-declared ``Sharpe no worse''
condition. The DEV placebo evidence (rank-based $p\approx0.024$) stands, but
the sealed window did not reproduce the Sharpe-neutral reduction.

\item \textbf{B7 and the selection lesson.} B7 again beat Config A primary on
growth ($+0.1079$ against $+0.0847$), reported here and in the abstract as
registered. But its mirror control B8, the identical rule on the other asset
class, scored $+0.1723$ with a 42.8\% drawdown, the best growth of any book. On
DEV, B7 won and B8 lost ($+0.2934$ against $+0.0892$); on the sealed window
they swapped. That reversal is the registered evidence that hindsight
asset-class books are regime lottery tickets, not a benchmark the method should
be required to beat---while noting that the method beat neither of them on
growth here.

\item \textbf{The registered prediction held.} Growth shrank materially from
the full-sample DEV anchor ($0.2813\to0.0847$, about 30\% of the development
value). The registered marginal standard error ($\pm0.14$) overstates the
uncertainty of highly correlated book-against-book differences, which is why
we now report paired intervals; they confirm that the window cannot sign most
comparisons (item 2).

\item \textbf{The registered risk-adjusted defence also failed.} On the sealed
window both configurations rank last of the eleven entries in
Tables~\ref{tab:l1} and~\ref{tab:l2} on both Sharpe (0.45 and 0.42 against
0.47--0.90) and Calmar (0.33 and 0.30 against 0.38--0.74), and Config A's
maximum drawdown (36.6\%) and Ulcer index (0.162) are worse than every passive
book B1--B6. The DEV separation on which Section~\ref{sec:threat} rests did not
transfer; we report that as a failure of the registered defence, not only of
the growth claim.

\item \textbf{A reporting error we caught, and what remains of it.} The
development records carried DEV turnover as 5.7$\times$, a full-sample average;
the DEV-window value is 14.1$\times$ (Config A) and 15.1$\times$ (Config B),
which reconciles with the cost sensitivity of Section~\ref{sec:robust}.
Sealed-window turnover of 16.0$\times$ is therefore a 13\% rise, not the
tripling an earlier draft reported. The cost half survives: at 50\,bps Config A
returns $+1.3$\%/yr and Config B is negative, so the development window's claim
of surviving a $10\times$ cost increase does not transfer.
\end{enumerate}

\subsection{How much of the fall is the environment?}
\label{sec:attrib}

Inverting $g=Lm-\tfrac12L^2s^2$ at $L=0.75$ on the harness-capped SPY book B1
implies an arithmetic SPY mean of 0.176 on DEV and 0.096 on the sealed window, a
factor of 0.55 (the DEV value returns uncapped SPY's 0.1594, as it should).
Scaling Config A's DEV equity contribution ($+0.237$ of its $+0.312$ arithmetic
total, Table~\ref{tab:perasset}) by that factor, setting its commodity
contribution to zero (the realized sealed-window figure is $+0.015$, against
$+0.075$ on DEV), and charging the higher variance drag (volatility
$23.4\%\to26.4\%$) predicts $+0.095$ against the realized $+0.085$. Most of the
fall from the development value is therefore environmental. This is arithmetic
on committed numbers; no configuration is scored and nothing is re-run. It does
not explain the shortfall against the passive books, which faced the same
environment; Section~\ref{sec:split} takes up the year in which that shortfall
arose.

\subsection{Decomposition, costs and financing}
\label{sec:lbdetail}

These are the registered diagnostics, computed by a companion script that
reuses the validated harness replica, because the DEV diagnostics script is
hard-wired to the development window; the script and its verbatim output are
committed.

\begin{table}[t]
\centering
\begin{minipage}[t]{0.40\linewidth}
\centering
\captionof{table}{Sealed-window per-year net log growth (raw, not annualised),
primary variants.}
\label{tab:l3}
\small
\begin{tabular}{@{}lcc@{}}
\toprule
year & Config A & Config B \\
\midrule
2022 & $-$0.288 & $-$0.273 \\
2023 & +0.368 & +0.314 \\
2024 (to Sep 20) & +0.149 & +0.149 \\
\bottomrule
\end{tabular}
\end{minipage}\hfill
\begin{minipage}[t]{0.56\linewidth}
\centering
\captionof{table}{Sealed-window cost sensitivity and financing overlay
(annualised net log growth).}
\label{tab:l4}
\small
\begin{tabular}{@{}lccccc@{}}
\toprule
cost (bps) & 0 & 5 & 10 & 20 & 50 \\
\midrule
Config A & +0.093 & +0.085 & +0.077 & +0.061 & +0.013 \\
Config B & +0.079 & +0.070 & +0.061 & +0.044 & $-$0.009 \\
\midrule
financing & 0\% & 1\% & 2\% & 4\% & \\
\midrule
Config A & +0.085 & +0.075 & +0.066 & +0.047 & \\
Config B & +0.070 & +0.062 & +0.055 & +0.039 & \\
\bottomrule
\end{tabular}
\end{minipage}
\end{table}

The sealed window is one bad year followed by two good ones
(Table~\ref{tab:l3}): neither configuration was stopped out, and the 2022 loss
sits inside the maximum drawdown already reported. \emph{Per asset} (Config A:
annualised net contribution / mean $|$position$|$ / coverage), every asset
contributed positively except USO ($-0.036$), with GLD the largest ($+0.033$);
coverage spans 0.698 (QQQ) to 0.834 (USO) against nominal 0.75---the energy
contracts were over-covered, the equity sleeve slightly under. \emph{Monthly
distribution} ($n=33$): Config A mean $+0.97$\%, standard deviation 7.5\%,
minimum $-13.0$\%, maximum $+16.5$\%, 48\% positive months, skew about zero;
Config B is tighter (standard deviation 7.0\%, minimum $-11.5$\%, 55\%
positive, positive skew), so the dial did reshape the distribution even though
its Sharpe criterion failed. At a favourable large-account financing rate of
about 4\% (typical retail margin runs far higher), Config A's sealed-window
growth is about $+4.7$\%/yr, below every unlevered passive book
(Table~\ref{tab:l4}). The strategy as configured is not investable, and we do
not recommend trading it.

\section{The coverage split: a post-hoc diagnostic}
\label{sec:split}

\begin{figure}[t]
\centering
\includegraphics[width=\linewidth]{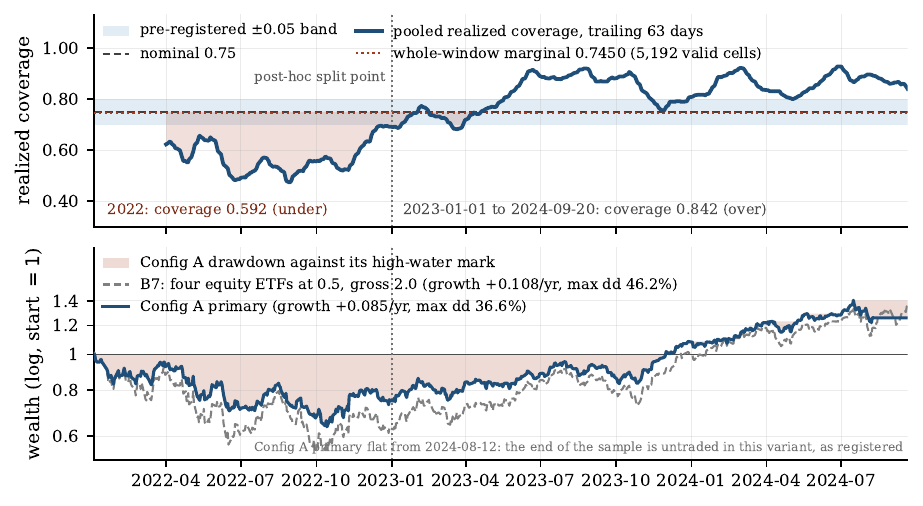}
\caption{The sealed window on one time axis. \emph{Top:} trailing 63-day
realized conformal coverage, pooled across the eight assets, against the
nominal 0.75 and the pre-registered $\pm0.05$ band; the whole-window marginal
coverage is 0.7450 over 5{,}192 valid cells. \emph{Bottom:} wealth of Config A
(primary) on a log scale, shaded against its high-water mark, with the
registered four-equity book B7. The under-covered stretch and the drawdown
occupy the same months, and B7 falls with Config A.}
\label{fig:split}
\end{figure}

\textbf{Status.} Figure~\ref{fig:split} and Table~\ref{tab:split} split the
sealed window at the year boundary. Nothing is re-scored: the split
re-aggregates the committed per-cell series of the same frozen artifact, which
reproduce Table~\ref{tab:l1} before they are split
(\texttt{paper/coverage\_split.py}). The split is post hoc twice over: the split
point was chosen after seeing the trailing-coverage figure, and the
registration explicitly forbade \emph{reporting a favourable sub-period}. We
report it because it is unfavourable; both halves are stated, neither is
promoted, and no registered number changes. That leaves a tension we do not
resolve: the most informative observation in the paper sits outside its
confirmatory protocol. We treat it as a hypothesis generated by the sealed
window, for a registered follow-up to test.

\begin{table}[t]
\centering
\caption{The sealed window split at the year boundary (Config A, primary).
Coverage with a 95\% block-bootstrap interval (63-day blocks, each half
resampled within itself) that ignores the post-hoc choice of split point;
growth is net log growth over the subperiod, raw and annualised; cap: share of
traded days on which pre-cap gross exceeded 2.0; equity: mean net weight of the
four equity ETFs in the held book.}
\label{tab:split}
\small
\setlength{\tabcolsep}{3.6pt}
\begin{tabular}{@{}lccccccc@{}}
\toprule
subperiod & coverage & cells & growth & ann. & cap & pre-cap & equity \\
\midrule
2022 & \textbf{0.592} \ci{0.53}{0.65} & 2{,}008 & \textbf{$-$0.288} & $-$28.9\% & 100\% & 4.50 & +1.32 \\
2023-01-01 to 2024-09-20 & \textbf{0.842} \ci{0.80}{0.88} & 3{,}184 & \textbf{+0.517} & +30.2\% & 99.5\% & 3.72 & +1.35 \\
whole window & 0.745 \ci{0.66}{0.82} & 5{,}192 & +0.230 & +8.5\% & 99.7\% & 4.02 & +1.34 \\
\midrule
DEV, same artifact & & & & & 99.9\% & 4.38 & +1.25 \\
\bottomrule
\end{tabular}
\end{table}

\textbf{What it shows.} The marginal 0.745 is not the coverage of a
well-calibrated window. It averages a 2022 in which realized coverage ran
0.592, nearly sixteen points below nominal and far outside the registered band,
and a 2023--24 in which it ran 0.842, nine points above. The gap of 0.25 is
large against block-bootstrap standard errors of 0.033 and 0.022 for the halves
(an interval for the gap of $[+0.17,+0.33]$), though two caveats apply: no
interval here accounts for the choice of split point, and with 63-day blocks a
251-day half holds only four blocks, so its standard error is itself rough (21-
and 126-day blocks give 0.038 and 0.030 for 2022). And the under-covered year
holds the entire loss: $-0.288$ of net log growth in 2022 against $+0.517$ over
the remaining 432 days, or $-28.9$\%/yr and $+30.2$\%/yr annualised. Both
rates are stated because neither is a rate: each covers a single sub-year
regime, each is a post-hoc slice, and the registered figure for this window is
the $+8.5$\%/yr of Table~\ref{tab:l1}.

\textbf{What it does not show.} The tempting reading is ``intervals too narrow,
so positions too large''. It does not survive the structural fact of
Section~\ref{sec:data}. The gross cap bound on all 251 trading days of 2022
(average pre-cap gross 4.50; Table~\ref{tab:split}), so the book held its
maximum exposure throughout and narrow intervals could not make it larger. Even
a uniform widening sufficient to restore nominal coverage---a factor of about
1.39 if 2022's residuals were roughly Gaussian, since coverage 0.592
corresponds to a half-width of 0.83 standard deviations against the 1.15 that
0.75 requires---would have divided raw Kelly weights by about 1.94 and left
average pre-cap gross above 2.3 (the per-asset clip makes that a lower bound),
still over the cap. Under the cap the interval sets composition, and the book's
net equity weight in 2022 ($+1.32$ of its 2.0 gross) sat between its DEV level
($+1.25$) and its 2023--24 level ($+1.35$). A common cause fits at least as
well. Table~\ref{tab:env} shows what happened to the assets themselves: equity
volatility nearly doubled in 2022 while equities fell, so a deliberately slow
quantile under-covered and a persistently long-equity book lost, in step with
B7 (Figure~\ref{fig:split}); when volatility fell back in 2023--24, the same
slow quantile over-covered. Finer composition effects are not ruled out---over
the whole window the registered per-asset diagnostics show the equity intervals
slightly under-covering and the energy intervals over-covering
(Section~\ref{sec:lbdetail})---but separating them from the common cause
requires scoring a recalibrated book, which the registration forbids.

\begin{table}[t]
\centering
\caption{The environment, from prices alone: annualised realized volatility
and arithmetic mean daily return (annualised) of SPY, the equal-weight four
equity ETFs, and the equal-weight eight ETFs.}
\label{tab:env}
\small
\begin{tabular}{@{}lcccccc@{}}
\toprule
 & \multicolumn{2}{c}{SPY} & \multicolumn{2}{c}{equity, 4 ETFs} & \multicolumn{2}{c}{all 8 ETFs} \\
period & vol & mean & vol & mean & vol & mean \\
\midrule
DEV 2016--2021 & 18.0\% & +17.6\% & 19.1\% & +18.4\% & 15.6\% & +13.1\% \\
2021 & 13.0\% & +26.1\% & 13.9\% & +23.8\% & 13.3\% & +21.7\% \\
2022 & 24.2\% & $-$17.2\% & 24.8\% & $-$17.0\% & 19.0\% & $-$0.7\% \\
2023-01-01 to 2024-09-20 & 12.9\% & +25.2\% & 13.5\% & +23.8\% & 11.8\% & +17.2\% \\
\bottomrule
\end{tabular}
\end{table}

\textbf{What survives.} Marginal coverage, which the registration called its
sharpest test, passed a window containing a sixteen-point conditional failure.
Here the binding cap kept that failure away from the book's size; without a
cap, a scale understated by a factor of about 1.4 would have meant positions
nearly twice too large for a year. That marginal validity does not constrain
conditional validity is a theorem~\citep{vovk2012,barber2021limits}; that it
can hide an error of this size in an interval used as a scale is one
observation. Together they motivate, and do not establish, registering
conditional coverage---by period, by volatility regime and by interval-width
quantile, for which methods with group-conditional guarantees
exist~\citep{gibbs2025}---as an endpoint for any rule that reads interval width
as a scale.

\textbf{One reading reconciles the two windows.} Under a binding cap only the
cross-section of widths matters, so a slow estimator's lag in \emph{level}---the
source of 2022's under-coverage---costs little, while its lower variance in the
\emph{ratios} helps. That is consistent with the development-window pattern
(slower was never worse in point estimate) and with the sealed window (a large
level error that did not reach the book's size). If it is right, slow
estimators should look worse off the cap, in exactly the regime shifts where
their lag is largest; that is the first test of the follow-up registration
(Section~\ref{sec:limits}).

\section{Limitations}
\label{sec:limits}

\begin{enumerate}[leftmargin=*]
\item \textbf{The development-window evidence is cap-conditional and untested
off the cap.} Every row of Tables~\ref{tab:sigma} and~\ref{tab:adaptive}---including the
2.1\,pp/yr the conformal quantile takes from the plug-in standard deviation, and
with it the tail-sensitivity ordering---was measured with the gross cap binding
on 97.7\% of days at the harness's \texttt{MAX\_GROSS = 2.0}. That is not a side
condition: Section~\ref{sec:data} derives the candidate mechanism from it. Below
the Kelly optimum, an estimator that understates $\sigma$ on the fat-tailed
assets keeps weight on them, and the book is rewarded for it only because the
cap holds it on the rising part of the growth-versus-exposure curve. With the
cap slack---at gross 1.0, or at 3.0 and 4.5, above the pre-cap average of 4.28,
where the book would sit near or past its own Kelly peak---that reward weakens
and, past the peak, reverses sign. The inertness of $\kappa$, the empirical
flatness of $\alpha$, the failure of inverse-volatility tilting, the
irrelevance of correlation and the impotence of constant de-levering all follow
from the same fact. Whether the ordering compresses, survives or inverts is
unknown, and it cannot be resolved here: the registration forbids scoring any
new configuration on any window after unsealing, and a gross sweep is a new
configuration on every window it touches. Until a registered gross-cap sweep is
run, the design evidence should be read as conditional on a binding cap of 2.0.
\item \textbf{Evidentiary status.} The paper's most informative observation
was not registered (Section~\ref{sec:split}). Its development comparisons are
exploratory, selected from a search of roughly 200 configurations, and 11 of
the 13 we could re-derive have bootstrap intervals that include zero; those
intervals are not adjusted for the search. Disclosure and the sealed window
mitigate the search; they do not cure it.
\item \textbf{One universe, one snapshot, one development window, one
2.7-year sealed window.} External validity is untested; a second universe was
forbidden by the frozen harness and is a high-value follow-up.
\item \textbf{The per-asset cap sits at its own optimum by luck.}
Section~\ref{sec:xsec}'s winsorisation dial optimises at $c\approx0.75$--$0.90$
and the harness fixes $c=0.75$ exogenously; at 0.30 or 2.0 the development
headline would be 4--8\,pp worse.
\item \textbf{2020 is a large fraction of the development result.} Config A's
2020 growth is $+0.667$ against a six-year mean of 0.285; the second half of DEV
is $2.6\times$ the first; monthly returns are strongly right-skewed (skew
$+1.49$, excess kurtosis $+7.65$) with a single $+36.5$\% month.
\item \textbf{Correlation is ignored.} Empirically inert under the binding cap
(Section~\ref{sec:corr}), but the per-asset formulation has no covariance term.
\item \textbf{Coverage is checked, not guaranteed.} Overlapping 21-day targets
break exchangeability, and three features of the construction each break the
argument on their own: the geometric shrinkage toward the expanding anchor, the
rolling calibration window, and the conformal-derived ensemble weights.
Marginal coverage is undamaged on both windows (0.7483 and 0.7450), and the
non-overlapping version behaves the same on DEV, but per-asset coverage spans
0.690--0.781 on DEV, and coverage by year spans 0.592--0.842 on the sealed
window.
\item \textbf{No live trading claim.} The harness charges 5\,bps of turnover and
nothing else; the financing overlay adds a borrowing estimate; neither
addresses market impact, borrow availability, margin-facility withdrawal risk
or execution. We are not recommending that this be traded.
\item \textbf{A zero risk-free rate is assumed throughout.} Sharpe and Calmar
are computed on raw rather than excess returns, and financing enters only as an
ex-post overlay. That is close to harmless on DEV, where short rates were near
zero, but it flatters every sealed-window risk-adjusted number, since cash paid
4--5\% over 2022--2024.
\item \textbf{The registered risk-adjusted defence did not survive the sealed
window.} Section~\ref{sec:threat} defends the method on risk-adjusted rather
than growth grounds, and that defence rests on DEV separation that did not
transfer (Section~\ref{sec:interpret}, item 6).
\end{enumerate}

\textbf{The follow-up registration}, in order: (1) a gross-cap sweep---slack,
binding, and above the pre-cap average; (2) conditional-coverage endpoints by
period, volatility regime and width quantile, with $q^2$-versus-realized-variance
diagnostics and a recalibrated-width control book that separates composition
from common cause; (3) a second universe; (4) risk-matched rather than
gross-matched $\sigma$ comparisons; (5) the specification with $z$ corrected;
(6) constant-scale and shuffled-width control books; and (7) validation of the
price series against an independent vendor---all with the bootstrap design of
Appendix~\ref{app:postreg} fixed in advance.

\section{Conclusion}
\label{sec:conclusion}

In one pre-registered eight-ETF experiment, marginal calibration transferred
out of sample and the economic value of the sizing rule did not: the capped
Kelly-inspired rule earned less than the passive benchmarks and ranked last
among them on Sharpe and Calmar, while its slowly varying conformal quantile
held near-nominal \emph{marginal} coverage. A post-hoc split shows that the
marginal figure averaged a sixteen-point under-coverage in 2022 with a later
over-coverage. The under-covered year held the entire loss, but the binding cap
rules out the simple oversizing story, and a regime that moved both coverage
and equity returns explains the co-location at least as well. The lesson we
draw is narrower than a causal one: an acceptance test on marginal coverage can
pass an interval-as-scale rule through a large conditional failure, so
conditional coverage should be registered alongside it, a recommendation this
experiment motivates rather than establishes.

The development-window evidence runs against the usual advice for conformal
prediction on time series: slow, unweighted, per-asset, tail-robust quantiles
out-grew every locally adaptive alternative tested, including the methods
designed for non-exchangeability, and a plug-in standard deviation by
2.1\,pp/yr at matched leverage; the same intervals, read as a pooled downside
break rate, cut development drawdown at better Sharpe and survived a
constant-leverage control and a timing placebo. That evidence is exploratory,
mostly within bootstrap noise contrast by contrast, and measured only under a
binding gross cap of 2.0; on the sealed window the drawdown control failed its
registered criterion. The candidate explanation---a scale estimator consumed by
a nonlinear sizing map is charged for its own estimation variance, while under a
binding cap its lag in level costs little---is a hypothesis for the follow-up
registration, not a finding of this one.

\appendix

\section{Experiment ledger and reproduction}
\label{app:ledger}

The full record is \texttt{results.tsv} (47 rows) and \texttt{NOTES.md}, both
committed on branch \texttt{autoresearch/\allowbreak jul30} of the project repository,
which---along with the frozen data snapshot, the pre-registration, the verbatim
sealed-window output and every post-registration script---is available from the
author on request. With the repository in hand:
\begin{verbatim}
git checkout best-leg3              # or paper-config-drawdown
uv run python train.py              # ~7 s
uv run python paper/diagnostics.py  # replays both tags
\end{verbatim}
\texttt{paper/diagnostics.py} loads each configuration from its git tag rather
than the working tree, re-implements the harness's position transform, and
\textbf{aborts unless that re-implementation reproduces
\texttt{prepare.evaluate\_positions} exactly}. It asserts
\texttt{DEV\_END == 2021-12-31} before doing anything and never touches sealed
data. A do-not-re-run list of every device tested and rejected closes each leg's
section in \texttt{NOTES.md}; Sections~\ref{sec:refute} and~\ref{sec:robust}
report the substantive entries.

\section{Post-registration analyses}
\label{app:postreg}

Every computation made after the single registered run is listed here, with
what it may and may not change. None scores a configuration for the first time,
none sets \texttt{FINAL\_EVAL}, and each asserts that the series it uses
reproduce the officially recorded numbers before reporting anything.

\begin{itemize}[leftmargin=*]
\item \texttt{paper/lockbox\_diagnostics.py}: the registered diagnostics
(per-year, per-asset, monthly, costs, financing; registration items 11--16).
\item \texttt{paper/lockbox\_figures.py}: the sealed-window wealth figure
(Figure~\ref{fig:lbwealth}); only registered books are drawn.
\item \texttt{paper/coverage\_split.py}: the year split of
Section~\ref{sec:split}. Post hoc, and in tension with the registration's
prohibition on sub-period reporting; reported because it is unfavourable.
\item \texttt{paper/icaif-camera-ready/make\_figure.py}: Figure~\ref{fig:split}.
\item \texttt{paper/postreg\_uncertainty.py}: the block-bootstrap intervals of
this paper, the gross-cap and equity-weight statistics of
Table~\ref{tab:split}, and the price-only environment of Table~\ref{tab:env}.
For the development contrasts it re-derives, from git, the daily series of 18
arms scored before unsealing and recorded in \texttt{results.tsv}, and asserts
each against its ledger value (to $10^{-6}$ for six-decimal entries, to
$5\times10^{-5}$ for four-decimal ones); the only configuration values it sets
select ablation arms whose results the ledger already records. It also
reproduces the paper's ``pre-cap gross 4.28, cap binding on 97.7\% of DEV
days'' from the tagged artifact.
\item The growth attribution of Section~\ref{sec:attrib} is arithmetic on
committed numbers.
\end{itemize}

\textbf{Bootstrap design.} Circular moving-block bootstrap; for paired growth
differences the statistic is $252\times$ the mean daily difference in
$\log(1+r)$, which equals the difference in reported growth exactly; for
coverage, days are resampled in blocks with all eight assets of a day kept
together, and the statistic is the pooled ratio of covered to valid cells.
10{,}000 resamples, fixed seed, common resampled days across development
contrasts, 95\% percentile intervals; block lengths of 21, 63 and 126 trading
days, with 63 (three forecast horizons) as the headline.

\begin{table}[t]
\centering
\caption{Development window: paired growth differences (pp/yr) with 95\%
moving-block-bootstrap intervals at three block lengths.}
\label{tab:bootdev}
\small
\begin{tabular}{@{}lcccc@{}}
\toprule
contrast & diff & $L=21$ & $L=63$ & $L=126$ \\
\midrule
Config A $-$ Config B & $+2.6$ & $[-1.8,+7.5]$ & $[-0.6,+6.4]$ & $[-0.2,+5.8]$ \\
$\sigma$, leg 2: conformal $-$ sd (matched) & $+2.1$ & $[-0.4,+4.7]$ & $[-0.5,+5.0]$ & $[-0.4,+4.6]$ \\
$\sigma$, leg 3: conformal $-$ sd & $+3.5$ & $[-0.1,+7.6]$ & $[-0.8,+9.3]$ & $[-0.9,+9.6]$ \\
$\sigma$, leg 3: conformal $-$ mean abs.\ dev. & $+1.3$ & $[-1.1,+4.0]$ & $[-1.5,+4.8]$ & $[-1.4,+5.0]$ \\
$\sigma$, leg 3: mean abs.\ dev.\ $-$ sd & $+2.2$ & $[+0.5,+4.1]$ & $[+0.1,+4.8]$ & $[+0.1,+4.9]$ \\
$\sigma$, leg 3: rolling $-$ frozen conformal & $+1.5$ & $[-1.9,+4.9]$ & $[-2.0,+5.1]$ & $[-2.2,+5.2]$ \\
$\sigma$, leg 3: frozen conformal $-$ rolling sd & $+1.9$ & $[-3.3,+7.5]$ & $[-4.3,+9.5]$ & $[-5.1,+10.7]$ \\
adaptive conformal inference $-$ base & $-1.6$ & $[-4.9,+1.6]$ & $[-5.1,+1.6]$ & $[-4.6,+0.9]$ \\
recency-weighted $-$ base & $-1.4$ & $[-4.2,+1.4]$ & $[-4.3,+1.5]$ & $[-3.7,+1.1]$ \\
asymmetric CQR-style $-$ base & $-1.6$ & $[-3.9,+0.5]$ & $[-3.6,+0.2]$ & $[-3.6,+0.2]$ \\
Mondrian (asset class) $-$ base & $-2.4$ & $[-6.2,+0.8]$ & $[-6.8,+1.1]$ & $[-6.9,+1.3]$ \\
pooled (all eight) $-$ base & $-2.5$ & $[-5.6,+0.2]$ & $[-6.6,+0.7]$ & $[-6.8,+0.7]$ \\
holding-period residual, $H_s=5$ $-$ base & $-0.7$ & $[-2.2,+0.7]$ & $[-2.0,+0.5]$ & $[-1.9,+0.3]$ \\
conformal predictive decision rule $-$ base & $-3.6$ & $[-7.1,-0.2]$ & $[-6.7,-0.4]$ & $[-6.2,-1.0]$ \\
Config A $-$ B4 (equal weight $2\times$) & $+7.2$ & $[-13.4,+30.9]$ & $[-12.3,+35.2]$ & $[-10.7,+34.3]$ \\
Config A $-$ B5 (inverse-vol $2\times$) & $+6.2$ & $[-10.2,+25.7]$ & $[-9.6,+28.9]$ & $[-8.3,+28.3]$ \\
Config A $-$ B7 (4 equity @ 0.5) & $-0.9$ & $[-18.4,+21.0]$ & $[-17.7,+22.2]$ & $[-16.4,+20.9]$ \\
\bottomrule
\end{tabular}
\end{table}

\begin{table}[t]
\centering
\caption{Sealed window (Config A primary): realized coverage and paired growth
differences against each registered book, with 95\% moving-block-bootstrap
intervals. Coverage intervals for the halves ignore the post-hoc choice of
split point.}
\label{tab:bootlb}
\small
\begin{tabular}{@{}lcccc@{}}
\toprule
quantity & value & $L=21$ & $L=63$ & $L=126$ \\
\midrule
coverage, whole window & 0.7450 & $[0.69,0.80]$ & $[0.66,0.82]$ & $[0.64,0.84]$ \\
coverage, 2022 & 0.5916 & $[0.52,0.67]$ & $[0.53,0.65]$ & $[0.54,0.65]$ \\
coverage, 2023-01-01 to 2024-09-20 & 0.8417 & $[0.80,0.89]$ & $[0.80,0.88]$ & $[0.81,0.87]$ \\
coverage gap, later minus 2022 & $+0.251$ & $[+0.16,+0.34]$ & $[+0.17,+0.33]$ & $[+0.19,+0.32]$ \\
\midrule
growth, Config A $-$ Config B (pp) & $+1.5$ & $[-5.5,+8.6]$ & $[-4.1,+6.9]$ & $[-4.1,+7.2]$ \\
growth, Config A $-$ B1 (pp) & $+2.2$ & $[-15.6,+20.8]$ & $[-15.0,+17.8]$ & $[-16.2,+19.1]$ \\
growth, Config A $-$ B2 (pp) & $-1.0$ & $[-21.2,+19.7]$ & $[-22.9,+19.0]$ & $[-25.5,+19.4]$ \\
growth, Config A $-$ B3 (pp) & $-0.2$ & $[-22.4,+22.4]$ & $[-22.0,+19.9]$ & $[-25.6,+20.8]$ \\
growth, Config A $-$ B4 (pp) & $-8.3$ & $[-28.1,+10.7]$ & $[-30.0,+10.8]$ & $[-31.0,+9.0]$ \\
growth, Config A $-$ B5 (pp) & $-7.3$ & $[-25.4,+10.0]$ & $[-28.0,+9.8]$ & $[-29.5,+8.6]$ \\
growth, Config A $-$ B6 (pp) & $-7.1$ & $[-24.9,+10.1]$ & $[-28.0,+10.0]$ & $[-29.9,+9.3]$ \\
growth, Config A $-$ B7 (pp) & $-2.3$ & $[-18.8,+14.8]$ & $[-16.4,+12.0]$ & $[-12.5,+8.7]$ \\
growth, Config A $-$ B8 (pp) & $-8.8$ & $[-52.3,+33.8]$ & $[-54.4,+33.6]$ & $[-56.2,+26.6]$ \\
growth, Config A $-$ B9 (pp) & $-2.3$ & $[-20.3,+16.4]$ & $[-18.1,+14.1]$ & $[-15.4,+11.7]$ \\
\bottomrule
\end{tabular}
\end{table}

\section{Additional development-window figures}
\label{app:figs}

\begin{figure}[H]
\centering
\includegraphics[width=0.80\linewidth]{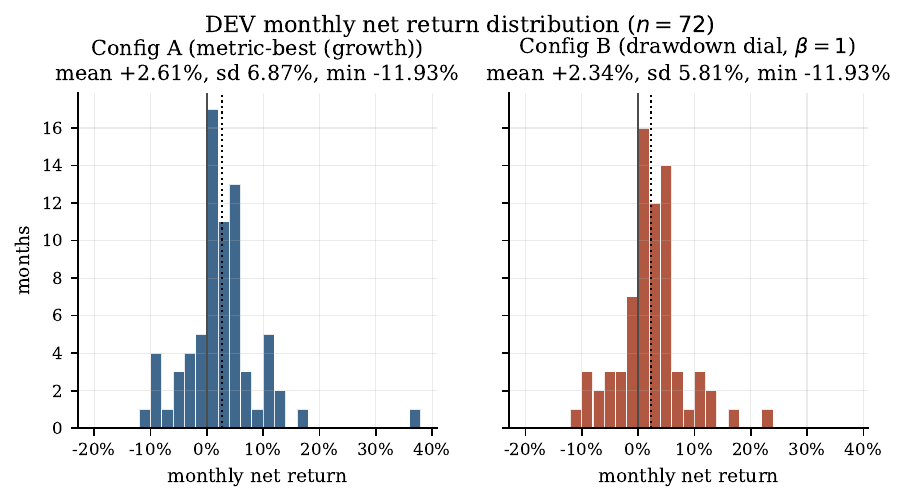}
\caption{DEV monthly net return distributions, Config A and Config B ($n=72$).}
\label{fig:monthly}
\end{figure}

\begin{figure}[H]
\centering
\includegraphics[width=0.86\linewidth]{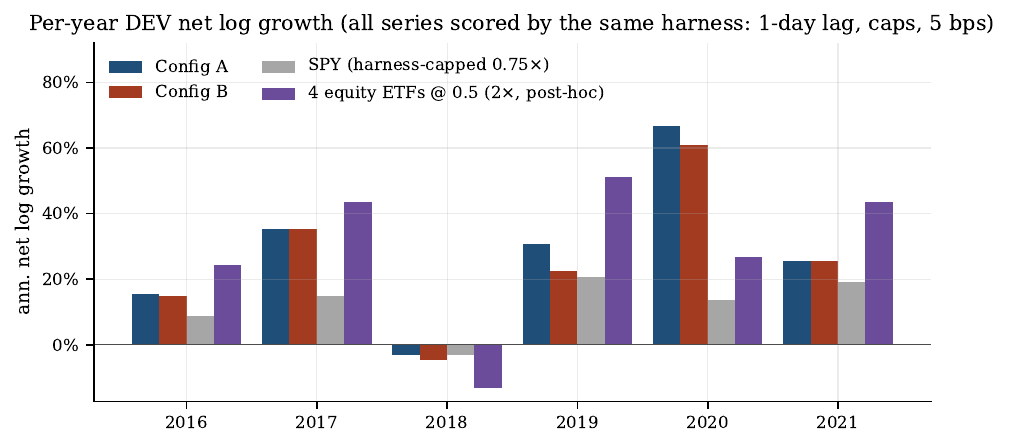}
\caption{Per-year DEV net log growth: Config A, Config B, harness-capped SPY,
and the post-hoc four-equity book B7, all scored by the same harness.}
\label{fig:peryear}
\end{figure}

\begin{figure}[H]
\centering
\includegraphics[width=0.80\linewidth]{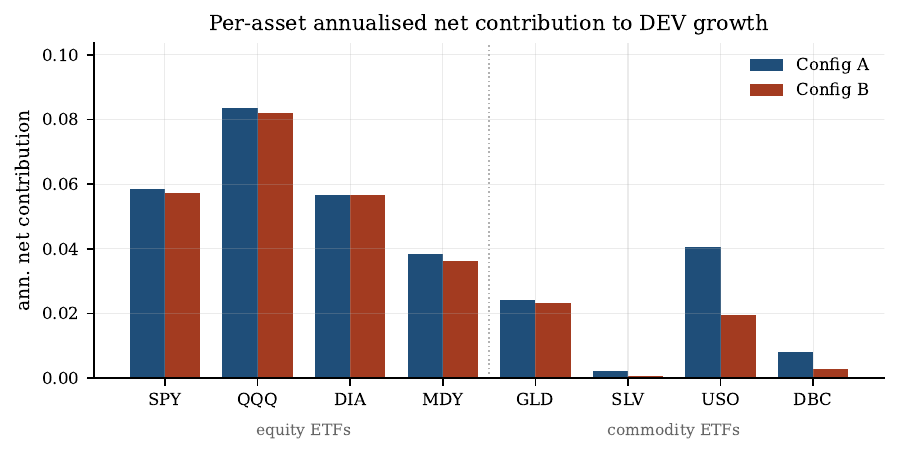}
\caption{Per-asset annualised net contribution to DEV growth, Config A and
Config B.}
\label{fig:perasset}
\end{figure}

\clearpage
\bibliographystyle{plainnat}
\bibliography{references}

@article{bailey2014dsr,
  author  = {Bailey, D. H. and L{\'o}pez de Prado, M.},
  title   = {The deflated {Sharpe} ratio: correcting for selection bias,
             backtest overfitting, and non-normality},
  journal = {The Journal of Portfolio Management},
  volume  = {40},
  number  = {5},
  pages   = {94--107},
  year    = {2014}
}

@article{bailey2017pbo,
  author  = {Bailey, D. H. and Borwein, J. M. and L{\'o}pez de Prado, M. and
             Zhu, Q. J.},
  title   = {The probability of backtest overfitting},
  journal = {Journal of Computational Finance},
  volume  = {20},
  number  = {4},
  year    = {2017},
  doi     = {10.21314/JCF.2016.322}
}

@article{barber2021limits,
  author  = {Barber, R. F. and Cand{\`e}s, E. J. and Ramdas, A. and
             Tibshirani, R. J.},
  title   = {The limits of distribution-free conditional predictive inference},
  journal = {Information and Inference: A Journal of the IMA},
  volume  = {10},
  number  = {2},
  pages   = {455--482},
  year    = {2021},
  doi     = {10.1093/imaiai/iaaa017}
}

@article{barber2023,
  author  = {Barber, R. F. and Cand{\`e}s, E. J. and Ramdas, A. and
             Tibshirani, R. J.},
  title   = {Conformal prediction beyond exchangeability},
  journal = {The Annals of Statistics},
  volume  = {51},
  number  = {2},
  pages   = {816--845},
  year    = {2023},
  doi     = {10.1214/23-AOS2276}
}

@article{browne1997,
  author  = {Browne, S.},
  title   = {Survival and growth with a liability: optimal portfolio
             strategies in continuous time},
  journal = {Mathematics of Operations Research},
  volume  = {22},
  number  = {2},
  pages   = {468--493},
  year    = {1997}
}

@misc{chan2024,
  author       = {Chan, J. S. and others},
  title        = {{MLE}-bench: evaluating machine learning agents on machine
                  learning engineering},
  year         = {2024},
  howpublished = {arXiv:2410.07095 (ICLR 2025)}
}

@inproceedings{gibbs2021,
  author    = {Gibbs, I. and Cand{\`e}s, E.},
  title     = {Adaptive conformal inference under distribution shift},
  booktitle = {Advances in Neural Information Processing Systems 34},
  year      = {2021}
}

@article{gibbs2025,
  author  = {Gibbs, I. and Cherian, J. J. and Cand{\`e}s, E. J.},
  title   = {Conformal prediction with conditional guarantees},
  journal = {Journal of the Royal Statistical Society Series B: Statistical
             Methodology},
  volume  = {87},
  number  = {4},
  pages   = {1100--1126},
  year    = {2025},
  doi     = {10.1093/jrsssb/qkaf008}
}

@article{grossman1993,
  author  = {Grossman, S. J. and Zhou, Z.},
  title   = {Optimal investment strategies for controlling drawdowns},
  journal = {Mathematical Finance},
  volume  = {3},
  number  = {3},
  pages   = {241--276},
  year    = {1993}
}

@article{jegadeesh1993,
  author  = {Jegadeesh, N. and Titman, S.},
  title   = {Returns to buying winners and selling losers: implications for
             stock market efficiency},
  journal = {The Journal of Finance},
  volume  = {48},
  number  = {1},
  pages   = {65--91},
  year    = {1993}
}

@inproceedings{jiahan2026,
  author    = {Jia, Y. and Han, B.},
  title     = {Portfolio selection with adaptive conformal prediction},
  booktitle = {Trends and Applications in Knowledge Discovery and Data
               Mining: PAKDD 2026 Workshops and Doctoral Consortium},
  series    = {Lecture Notes in Computer Science},
  volume    = {16603},
  pages     = {312--323},
  publisher = {Springer},
  address   = {Singapore},
  year      = {2026},
  doi       = {10.1007/978-981-92-2014-4_25}
}

@misc{karpathy2026,
  author       = {Karpathy, A.},
  title        = {autoresearch: {AI} agents running research on single-{GPU}
                  nanochat training automatically},
  year         = {2026},
  howpublished = {GitHub repository, MIT licence},
  url          = {https://github.com/karpathy/autoresearch}
}

@misc{kato2024,
  author       = {Kato, M.},
  title        = {Conformal predictive portfolio selection},
  year         = {2024},
  howpublished = {arXiv:2410.16333}
}

@article{kelly1956,
  author  = {Kelly, Jr., J. L.},
  title   = {A new interpretation of information rate},
  journal = {Bell System Technical Journal},
  volume  = {35},
  number  = {4},
  pages   = {917--926},
  year    = {1956}
}

@misc{lu2024,
  author       = {Lu, C. and Lu, C. and Lange, R. T. and Foerster, J. and
                  Clune, J. and Ha, D.},
  title        = {The {AI} {Scientist}: towards fully automated open-ended
                  scientific discovery},
  year         = {2024},
  howpublished = {arXiv:2408.06292}
}

@article{maclean2010,
  author  = {MacLean, L. C. and Thorp, E. O. and Ziemba, W. T.},
  title   = {Long-term capital growth: the good and bad properties of the
             {Kelly} and fractional {Kelly} capital growth criteria},
  journal = {Quantitative Finance},
  volume  = {10},
  number  = {7},
  pages   = {681--687},
  year    = {2010}
}

@book{maclean2011,
  editor    = {MacLean, L. C. and Thorp, E. O. and Ziemba, W. T.},
  title     = {The {Kelly} Capital Growth Investment Criterion: Theory and
               Practice},
  publisher = {World Scientific},
  year      = {2011}
}

@article{maclean1992,
  author  = {MacLean, L. C. and Ziemba, W. T. and Blazenko, G.},
  title   = {Growth versus security in dynamic investment analysis},
  journal = {Management Science},
  volume  = {38},
  number  = {11},
  pages   = {1562--1585},
  year    = {1992}
}

@misc{malik2024,
  author       = {{malik1641}},
  title        = {Stocks and {ETFs} Prices [data set; snapshot 2024-09-25;
                  prices sourced from {Yahoo Finance}]},
  year         = {2024},
  howpublished = {Kaggle},
  url          = {https://www.kaggle.com/datasets/malik1641/stocks-and-etfs-prices}
}

@misc{noguer2024,
  author       = {Noguer i Alonso, M.},
  title        = {Conformal portfolio optimization},
  year         = {2024},
  howpublished = {SSRN working paper 5011129}
}

@misc{noguer2024b,
  author       = {Noguer i Alonso, M.},
  title        = {Conformal prediction in finance},
  year         = {2024},
  howpublished = {SSRN working paper 4939336}
}

@article{phipson2010,
  author  = {Phipson, B. and Smyth, G. K.},
  title   = {Permutation p-values should never be zero: calculating exact
             p-values when permutations are randomly drawn},
  journal = {Statistical Applications in Genetics and Molecular Biology},
  volume  = {9},
  number  = {1},
  pages   = {Article 39},
  year    = {2010}
}

@misc{ray2026,
  author       = {Ray, R. D.},
  title        = {{ARTEMIS}: a neuro symbolic framework for economically
                  constrained market dynamics},
  year         = {2026},
  howpublished = {arXiv:2603.18107}
}

@inproceedings{romano2019,
  author    = {Romano, Y. and Patterson, E. and Cand{\`e}s, E. J.},
  title     = {Conformalized quantile regression},
  booktitle = {Advances in Neural Information Processing Systems 32},
  year      = {2019}
}

@misc{sunboyd2018,
  author       = {Sun, Q. and Boyd, S.},
  title        = {Distributional robust {Kelly} gambling: optimal strategy
                  under uncertainty in the long-run},
  year         = {2018},
  howpublished = {arXiv:1812.10371}
}

@inproceedings{vovk2018,
  author    = {Vovk, V. and Bendtsen, C.},
  title     = {Conformal predictive decision making},
  booktitle = {Proceedings of the Seventh Workshop on Conformal and
               Probabilistic Prediction and Applications},
  series    = {PMLR},
  volume    = {91},
  pages     = {52--62},
  year      = {2018}
}

@book{vovk2005,
  author    = {Vovk, V. and Gammerman, A. and Shafer, G.},
  title     = {Algorithmic Learning in a Random World},
  publisher = {Springer},
  note      = {Second edition 2022},
  year      = {2005}
}

@techreport{vovk2003,
  author      = {Vovk, V. and Lindsay, D. and Nouretdinov, I. and
                 Gammerman, A.},
  title       = {Mondrian confidence machine},
  institution = {Royal Holloway, University of London},
  type        = {On-line Compression Modelling project working paper},
  year        = {2003}
}

@inproceedings{vovk2012,
  author    = {Vovk, V.},
  title     = {Conditional validity of inductive conformal predictors},
  booktitle = {Proceedings of the Asian Conference on Machine Learning},
  series    = {PMLR},
  volume    = {25},
  pages     = {475--490},
  year      = {2012}
}

@article{vovk2019,
  author  = {Vovk, V. and Shen, J. and Manokhin, V. and Xie, M.},
  title   = {Nonparametric predictive distributions based on conformal
             prediction},
  journal = {Machine Learning},
  volume  = {108},
  number  = {3},
  pages   = {445--474},
  year    = {2019}
}

@misc{vovk2024,
  author       = {Vovk, V.},
  title        = {An optimality property of the {Bayes}--{Kelly} algorithm},
  year         = {2024},
  howpublished = {arXiv:2402.03035}
}

@inproceedings{xu2021,
  author    = {Xu, C. and Xie, Y.},
  title     = {Conformal prediction interval for dynamic time-series},
  booktitle = {Proceedings of the 38th International Conference on Machine
               Learning},
  series    = {PMLR},
  volume    = {139},
  pages     = {11559--11569},
  year      = {2021}
}

\end{document}